\documentclass[a4paper,usenatbib]{mnras}

\usepackage{graphicx}
\usepackage{amssymb}

\title[Five-year unWISE Coadds]{unWISE Coadds: The Five-year Data Set}

\author[Meisner et al.]{
A.~M. Meisner,$^{1}$\thanks{ameisner@noao.edu}
D. Lang,$^{2}$
E.~F. Schlafly$^{3, 4}$
and D.~J. Schlegel$^{3}$
\\
$^{1}$National Optical Astronomy Observatory, 950 N. Cherry Ave., Tucson, AZ 85719, USA \\
$^{2}$Perimeter Institute, 31 Caroline Street North, Waterloo, ON, N2L 2Y5, Canada \\
$^{3}$Lawrence Berkeley National Laboratory, Berkeley, CA, 94720, USA \\
$^{4}$Lawrence Livermore National Laboratory, Livermore, CA, 94551, USA \\
}

\pubyear{2019}

\begin{document}
\label{firstpage}
\pagerange{\pageref{firstpage}--\pageref{lastpage}}
\maketitle

\begin{abstract}
We present full-sky coadded maps created by uniformly combining the first five years of Wide-field Infrared Survey Explorer (WISE) and NEOWISE imaging at 3.4$\mu$m (W1) and 4.6$\mu$m (W2). By incorporating both pre-hibernation WISE exposures from 2010-2011 and the first four years (2013-2017) of post-hibernation exposures from the NEOWISE-Reactivation mission, we are able to provide W1/W2 coadds that span a 15$\times$ longer time baseline and are substantially deeper than the standard AllWISE data products. Our new five-year ``full-depth'' coadds are now the deepest ever all-sky maps at 3$-$5$\mu$m, permitting detection of sources $\sim$2$\times$  ($\sim$0.7 mag) fainter than AllWISE at 5$\sigma$ significance. We additionally present an updated set of ``time-resolved'' W1/W2 coadds, which separately stack each of $\sim$10 sky passes at each inertial sky location, enabling motion and variability measurements for faint infrared sources over a long $\sim$7.5 year time baseline. We highlight new processing improvements relative to our previous ``unWISE'' coadd releases, focusing on astrometric calibration and artifact flagging. The deep WISE stacks presented here are already being used to perform target selection for the Dark Energy Spectroscopic Instrument (DESI), and our full-sky coadded WISE/NEOWISE products will be key precursor data sets for upcoming wide-field infrared missions including SPHEREx and NEOCam.
\end{abstract}

\begin{keywords}

atlases --- infrared: general --- methods: data analysis --- surveys  --- techniques: image processing
\end{keywords}

\section{Introduction}
\label{sec:intro}
In the coming years, an exciting array of space-based infrared astronomy missions are expected to launch, including SPHEREx \citep{spherex}, JWST \citep{jwst}, Euclid \citep{euclid}, WFIRST \citep{wfirst} and NEOCam \citep{neocam}. For the purposes of forecasting/simulating the data sets that will be obtained by these projects, and to select optimal targets for their pointed observations, it is critical to fully process and analyze existing all-sky infrared imaging.

Launched in late 2009, the Wide-field Infrared Survey Explorer \citep[WISE;][]{wright10} is an unrivaled source of full-sky imaging in the mid-infrared. WISE has now provided the best ever all-sky sets of images over the 3$-$22$\mu$m wavelength range, in four broad channels labeled W1 (3.4$\mu$m), W2 (4.6$\mu$m), W3 (12$\mu$m) and W4 (22$\mu$m). WISE's sensitivity is orders of magnitude better than that of its predecessor, the Infrared Astronomical Satellite \citep[IRAS;][]{iras}.

The WISE satellite's original objectives were primarily to study the most infrared-luminous galaxies in the Universe \citep[e.g.,][]{tsai15} and identify extremely cold members of the Sun's local Galactic neighborhood \citep[e.g.,][]{kirkpatrick11}. However, by now, more than 80\% of WISE data at W1 and W2 has been acquired as part of the NEOWISE mission extensions \citep{neowise, neowiser}. NEOWISE is tasked with finding and characterizing asteroids, so the mission itself does not produce coadded data products of the sort that would maximize the value of its vast archive of raw imaging for Galactic and extragalactic astrophysics.

As a result, we have been undertaking a wide-ranging archival analysis program to repurpose the entire publicly available NEOWISE data set for astrophysics beyond the solar system. This effort begins by uniformly stacking all WISE and NEOWISE single-exposure images, with our data products referred to as ``unWISE'' coadds \citep{lang14}. Our unWISE coadds have already provided the deepest ever full-sky maps \citep{fulldepth_neo1, fulldepth_neo2, fulldepth_neo3}, source catalogs \citep{unwise_catalog}, and proper motion measurements\footnote{Delivered via the CatWISE catalog \citep{catwise_preliminary} at \url{https://catwise.github.io/}.} at 3$-$5$\mu$m. The Dark Energy Spectroscopic Instrument \citep[DESI;][]{desi, desi_part1, desi_part2} pre-imaging surveys have also published forced photometry of our unWISE coadds for more than 1 billion optically selected sources over $\sim$1/3 of the sky \citep{dey_overview, lang14b}.

Still, yet more single-exposure NEOWISE data remain to be incorporated into our analysis. Here, we update our unWISE coadd data set to include the fourth year of NEOWISE-Reactivation \citep[NEOWISE-R;][]{neowiser} exposures, which were gathered throughout 2017. In combination with existing WISE/NEOWISE data, this allows us to create yet deeper full-sky maps at 3$-$5$\mu$m and extend the time baseline to 7.5 years for proper motion and variability measurements based on epochal unWISE coadds \citep{tr_neo2, tr_neo3}. For this new ``five-year'' unWISE coadd data release, we have also implemented a number of processing improvements, particularly with regard to astrometric calibration and artifact flagging.

In $\S$\ref{sec:wise_mission} we briefly summarize the WISE/NEOWISE mission characteristics and timeline. In $\S$\ref{sec:data} we describe the input data used to build our five-year unWISE coadds. In $\S$\ref{sec:unwise} we review key aspects of the unWISE data products in relation to those published by the WISE/NEOWISE teams. In $\S$\ref{sec:neo4} we provide details of the five-year unWISE coaddition process and results. In $\S$\ref{sec:astrometry} we describe our astrometric recalibration of the five-year unWISE coadds using Gaia DR2 \citep{gaia_dr2}. In $\S$\ref{sec:artifact_summary} we give a high-level summary of the improved artifact flagging we have implemented, with full details deferred to Appendix \ref{app:bitmask}. In $\S$\ref{sec:validation} we discuss the validation of our five-year unWISE coadds. In $\S$\ref{sec:dr} we describe the five-year unWISE coadd data release. We conclude in $\S$\ref{sec:conclusion}.

\section{WISE Overview}
\label{sec:wise_mission}

WISE is a 40 cm telescope aboard a satellite in a $\sim$95 minute low-Earth orbit. After launching in late 2009, WISE surveyed the entire sky in all four of its channels (W1-W4) during the first half of 2010. In the latter half of 2010, the reddest two channels (W3 and W4) became unusable due to cryogen depletion. Nevertheless, the WISE satellite continued surveying in W1 and W2 until early 2011 as part of the NEOWISE mission \citep{neowise}. In 2011 February, WISE was placed into hibernation for approximately 33 months, during which time it was not conducting astronomical observations. In 2013 December, WISE recommenced surveying in the available W1 and W2 bands, and has continued doing so ever since thanks to the NEOWISE-R mission extension \citep{neowiser}. The very high W1/W2 data quality has remained essentially unchanged throughout the 9+ year timespan over which observations have been acquired in these two bands. The WISE angular resolution in W1 and W2 is $\sim$6$''$. WISE scans in great circles near solar elongation of $90^{\circ}$, with a typical sky region being observed over a $\gtrsim$1 day time period (referred to as a ``visit'') once every six months. During one such visit to a given sky location, $\gtrsim$12 single-exposure images per band are acquired. Very close to the ecliptic poles, the WISE time coverage becomes essentially continuous, with one exposure available during nearly every $\sim$95 minute orbit.

\section{WISE/NEOWISE Data}
\label{sec:data}
The NEOWISE-R mission publicly releases single-exposure images and catalogs, but no coadded data products. Therefore, the W1/W2 calibrated single-exposure (``L1b'') images provided by the WISE/NEOWISE teams represent the starting point of our unWISE coaddition processing. We begin by downloading all L1b images from the first five years of WISE observations: 13 months of pre-hibernation W1/W2 imaging plus the first four annual NEOWISE-R L1b data releases \citep{cutri15}. These data have acquisition dates spanning from 2010 January 7 to 2017 December 13. This full set of archival W1/W2 imaging constitutes a large volume of data: $\sim$13.1 million single-exposure images in each of two bands, totaling 175 terabytes. We note that a fifth year of post-reactivation W1/W2 single-exposure images recently became available on 2019 April 11 as part of the fifth annual NEOWISE-R data release, but we have not yet incorporated this most recent year of imaging into our unWISE coadds --- doing so will be the subject of future work.

\section{unWISE Overview}

\label{sec:unwise}
There are two lines of coadded full-sky WISE data products available: the AllWISE Source Catalog and Atlas stacks provided by the WISE/NEOWISE team \citep{cutri13}, and our unWISE coadds/catalogs. A major limitation of AllWISE for the W1/W2 channels is that it only incorporates the $\sim$1 year of pre-hibernation WISE imaging, which now represents just a small fraction of the presently available data in these two bands. In contrast, unWISE has been systematically reprocessing all W1/W2 data, both pre-hibernation and post-reactivation, as NEOWISE-R continues issuing its annual single-exposure releases \citep{fulldepth_neo1, fulldepth_neo2, tr_neo2, fulldepth_neo3, tr_neo3}. Prior to the present work, the most recent set of unWISE coadds incorporates a total of four years of single-exposure W1/W2 data (pre-hibernation imaging plus the first three years of post-reactivation exposures).

Another important point of contrast is that the AllWISE Atlas stacks are intended to be used primarily for performing source detection, and have therefore been in effect ``blurred'' by the WISE PSF. On the other hand, unWISE uses Lanczos interpolation when reprojecting the single-exposure L1b images in order to retain the native WISE resolution; this yields coadds that are unblurred and optimized for forced photometry \citep{lang14}.

unWISE coadds come in two flavors: ``full-depth'' and ``time-resolved''. The full-depth unWISE coadds simply stack together all available W1/W2 single-exposure images at each sky location to produce the deepest possible static sky maps. The time-resolved unWISE coadds bin the exposures at each sky location into a series of six-monthly visits, with typically $\gtrsim$12 exposures stacked together per band per visit\footnote{Full details of this time-slicing procedure are provided in $\S$3.2 of \cite{tr_neo2}.}. The time-resolved unWISE coadds therefore provide a means of measuring long-timescale ($\sim$6 months to $\sim$8 years) variability and motion for faint sources far below the single-exposure detection limit \citep[e.g.,][]{backyard_worlds, j1935, ross2018, stern2018}, with the effects of fast transients such as cosmic rays and satellite streaks dramatically suppressed. Because the present work incorporates five years of W1/W2 data, our outputs typically include 10 coadded epochs per band at each sky location, spanning a 7.5 year time baseline.

unWISE coaddition reprojects onto a set of 18,240 1.56$^{\circ}$$\times$1.56$^{\circ}$ `tile' footprints each identified by a unique string \verb|coadd_id| value that encodes the location of the tile center. For example, the unWISE tile centered at ($\alpha$, $\delta$) = (149.7485$^{\circ}$, 1.5144$^{\circ}$) has \verb|coadd_id| = 1497p015. The unWISE tile centers and orientations are chosen to match those of the AllWISE Atlas stacks. unWISE coadds adopt the native W1/W2 single-exposure pixel scale of $2.75''$/pixel, making each unWISE coadd image 2048$\times$2048 pixels in size. The unWISE coadd images are in units of Vega nanomaggies\footnote{A source with total flux of 1 nanomaggie has a magnitude of 22.5.}.

The ``unWISE Catalog'' \citep{unwise_catalog} is a WISE-selected full-sky source catalog created by modeling the five-year full-depth unWISE coadds presented in this work. Because the unWISE Catalog benefits from $\sim$5$\times$ more W1/W2 exposure time than AllWISE, it is $\sim$2$\times$ deeper than AllWISE in W1 and W2, extracting a total of $>$2 billion unique sources over the entire sky. We use the unWISE Catalog astrometry and photometry on several occasions throughout this paper to characterize and validate our five-year unWISE coadds.

\section{Five-year unWISE Coaddition}
\label{sec:neo4}
\subsection{Photometric calibration}

\begin{figure*}
        \includegraphics[width=7.0in]{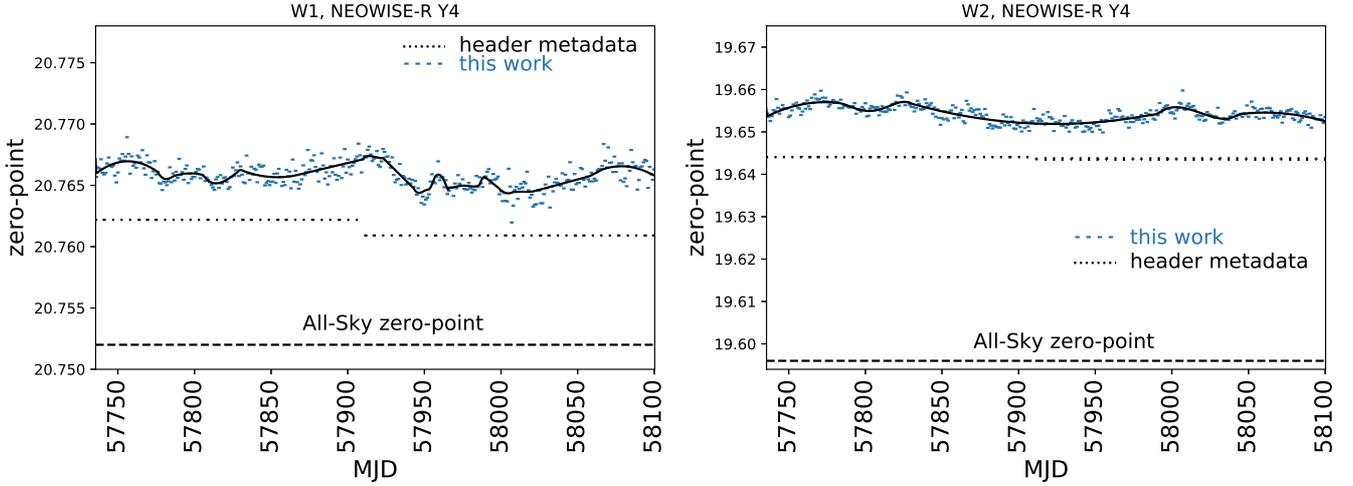}
    \caption{Our measured photometric zero-points for the fourth year of NEOWISE-R data, computed via the procedure detailed in $\S$4 of \citet{fulldepth_neo1}. Left: W1. Right: W2. Blue dashes indicate our per-day zero-point measurements. During coaddition, we interpolate off of each band's sequence of per-day measurements using smooth functions shown as solid black lines. Dotted black lines show the \texttt{MAGZP} zero-point values extracted from the L1b headers. Dashed black lines near the bottom of each panel show the All-Sky phase zero-point values.}
    \label{fig:zp}
\end{figure*}

Before proceeding to coadd the five years of W1/W2 L1b images downloaded, we begin by computing a custom photometric zero-point for each exposure so as to place all exposures on the most uniform possible relative photometric calibration. Per-exposure zero-points are needed to specify the multiplicative rescaling of each L1b image during unWISE coaddition. In brief, we measure the zero-point variation in each band in day-long intervals using repeat observations of stars near the ecliptic poles, which are observed on a frequent basis thanks to the WISE scan strategy. We find that this procedure captures subtle zero-point variations with time that are not reflected in the per-exposure zero-point provided by the \verb|MAGZP| keyword in each exposure's L1b header. Full details of our relative photometric calibration methodology can be found in $\S$4 of \cite{fulldepth_neo1}. Figure \ref{fig:zp} shows the results of our photometric zero-point determination procedure as applied to the fourth year of NEOWISE-R exposures (i.e., those exposures newly incorporated into our coadds in this work).

\subsection{Overview of five-year coaddition results}

We ran the unWISE coaddition code without substantive modifications relative to the most recent algorithmic updates described in \cite{fulldepth_neo2} and \cite{tr_neo2}, but now employing five years of W1/W2 single-exposure imaging, which is a larger quantity of raw data than previously used for any full-sky unWISE coadd processing. $\S$\ref{sec:fulldepth_numbers} and $\S$\ref{sec:tr_numbers} provide a high-level overview of the full-sky coaddition results.

\subsubsection{Five-year full-depth unWISE coadds}
\label{sec:fulldepth_numbers}

As described in $\S$\ref{sec:unwise}, the `full-depth' unWISE coadds simply stack all available frames together in each band at each sky location, resulting in the deepest possible static sky maps. Because this work incorporates five years of W1/W2 exposures, our new full-depth coadds now constitute the deepest ever full-sky maps at W1 and W2, as evidenced by the integer frame coverage (or equivalently the total exposure time) and further depth validations discussed in $\S$\ref{sec:validation}. Over the entire sky, the mean integer frame coverage is 178 (177) exposures per sky position in W1 (W2), $\sim$5$\times$ higher than that of the AllWISE Atlas mosaics. The minimum integer frame coverage is 65 (60) exposures in W1 (W2) --- no empty ``holes'' exist anywhere on the sky. The maximum integer frame coverage is 24,563 (24,514) exposures in W1 (W2), close to the north ecliptic pole. We did not specially discard any input frames near the ecliptic poles even though doing so would have been a computational convenience, as these regions are unusually deep and require hundreds of gigabytes of memory to coadd with our unWISE pipeline.

\subsubsection{Five-year time-resolved unWISE coadds}
\label{sec:tr_numbers}

\begin{figure}
        \includegraphics[width=3.3in]{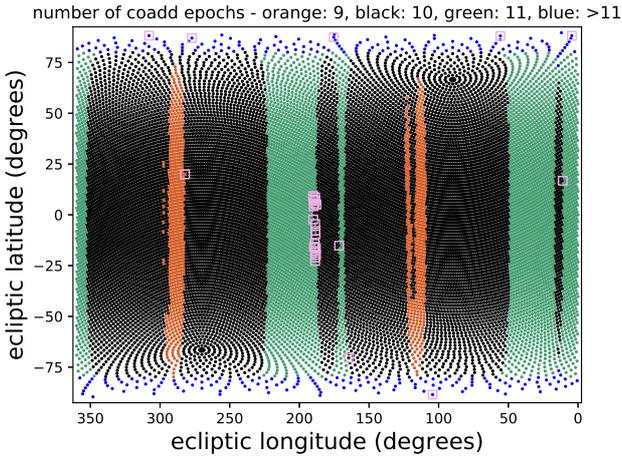}
    \caption{Full-sky map of the number of time-resolved unWISE coadd epochs per \texttt{coadd\_id} footprint, in W1. Pink squares indicate tiles with a differing number of coadd epochs in W1 versus W2. Usually there are ten epochal coadds per band per \texttt{coadd\_id} in the five-year unWISE data set (black points). 
    Orange regions missed one visit due to the April 2014 WISE command timing anomaly, and as a result have only nine epochal coadds per band.
    Green regions benefitted from coverage during the partial third pre-hibernation sky pass in early 2011, and therefore have an extra eleventh epoch. The ecliptic poles ($|\beta|$ $>$ 80$^{\circ}$, blue) feature $>$11 coadd epochs per band due to the WISE scan strategy and alternative unWISE time-slicing adopted in these regions, splitting the available exposures into a series of $\sim$10 day intervals \citep[][$\S$3.2.2]{tr_neo2}.}
    \label{fig:num_epochs}
\end{figure}

Our most recent prior release of time-resolved unWISE coadds \citep{tr_neo3} incorporated four years of W1/W2 imaging, and typically offered 8 time-resolved coadd epochs per band at a given sky position spanning a 6.5 year baseline. By adding in a fifth year of W1/W2 exposures, we generally obtain two additional time-resolved coadd epochs per band per sky location, while simultaneously extending the time baseline to 7.5 years. In our five-year unWISE coadd data set, every \verb|coadd_id| tile footprint has at least 9 time-resolved coadd epochs available per band. A median of 10 time-resolved coadd epochs per band is available over the full sky. The maximum number of time-resolved coadd epochs per band is 185, close to the ecliptic poles. Figure \ref{fig:num_epochs} shows the spatial distribution of the number of coadd epochs per band. This distribution of the number of coadd epochs matches expectations based on the time periods during which WISE was/wasn't observing, the WISE survey geometry, and our time-slicing algorithm. Over the entire sky, the total number of single-band epochal coadds in the five-year unWISE data set is 387,915.

\section{Astrometric Recalibration to Gaia DR2}
\label{sec:astrometry}

\subsection{Full-depth unWISE coadd astrometry}

As discussed in $\S$\ref{sec:unwise}, the unWISE coadds are optimized for forced photometry, in particular that performed to enable DESI's selection of luminous red galaxy and quasar targets as cosmological tracers \citep{desi_part1, dey_overview}. Because such forced photometry adopts source positions determined in the optical when measuring off of the full-depth unWISE coadds, it is critical that those coadds be accurately astrometrically calibrated, since any source mis-centering will lead to WISE flux underestimation. As a result, spatially coherent systematics in the full-depth unWISE coadd astrometry can imprint onto the DESI selection function, potentially biasing downstream power spectrum measurements. It is therefore very important to assess and optimize the quality of the full-depth unWISE coadd astrometry.

In previous work \citep{tr_neo2, tr_neo3}, we have presented a detailed characterization of the unWISE time-resolved coadd astrometry, along with refined world coordinate system (WCS) parameters designed to optimize the time-resolved coadds for proper motion studies. However, until now we have not provided such an astrometric analysis for our full-depth unWISE coadds, which are constructed by simply propagating the L1b WCS without modification. Making use of high-fidelity astrometry from the recently released Gaia DR2 \citep{gaia_dr2}, we present an assessment of the five-year full-depth unWISE coadd astrometry and describe our corresponding creation of improved WCS solutions. To do so, we obtain centroids of sources in the five-year unWISE stacks from the full-sky unWISE Catalog \citep{unwise_catalog}, which provides completely independent astrometric measurements in each of the W1 and W2 bands. The unWISE Catalog astrometry is based on model fits, and for the purposes of this analysis is equivalent to simple flux-weighted centroiding, since the unWISE Catalog PSF models are constructed so as to always place the flux-weighted centroid at the origin.

\subsubsection{Computing per-coadd offsets relative to Gaia}
We begin our astrometry evaluation process by computing per-coadd median positional offsets relative to Gaia for each full-depth unWISE image. Each full-depth coadd identified by its unique (\verb|coadd_id|, band) pair is analyzed independently. We seek to compute median per-coadd offsets along the unWISE x and y pixel coordinate directions, and in RA and Dec. For each unWISE \verb|coadd_id| footprint, we gather the list of Gaia DR2 sources that fall within this footprint and have full astrometric solutions available. Without a full astrometric solution available, we would not be able to account for the motion of each calibrator source. The Gaia proper motion allows us to accurately propagate the Gaia coordinates to the mean epoch of each unWISE coadd before performing any unWISE versus Gaia comparisons. We use the unWISE \verb|-frames| single-exposure metadata table corresponding to each coadd to compute the mean MJD of the contributing frames. We then propagate Gaia positions to the mean epoch of each full-depth unWISE coadd based on the Gaia proper motions. We ignore parallax because the full-depth unWISE coadds average together equal amounts of imaging on opposite sides of the parallactic ellipse as a result of the WISE survey strategy. The typical mean epoch of the five-year full-depth unWISE coadds is MJD $\approx$ 56970 (year $\approx$ 2014.85), whereas Gaia DR2 natively reports positions at epoch 2015.5. 

\begin{figure*}
\begin{centering}
       \includegraphics[width=5in]{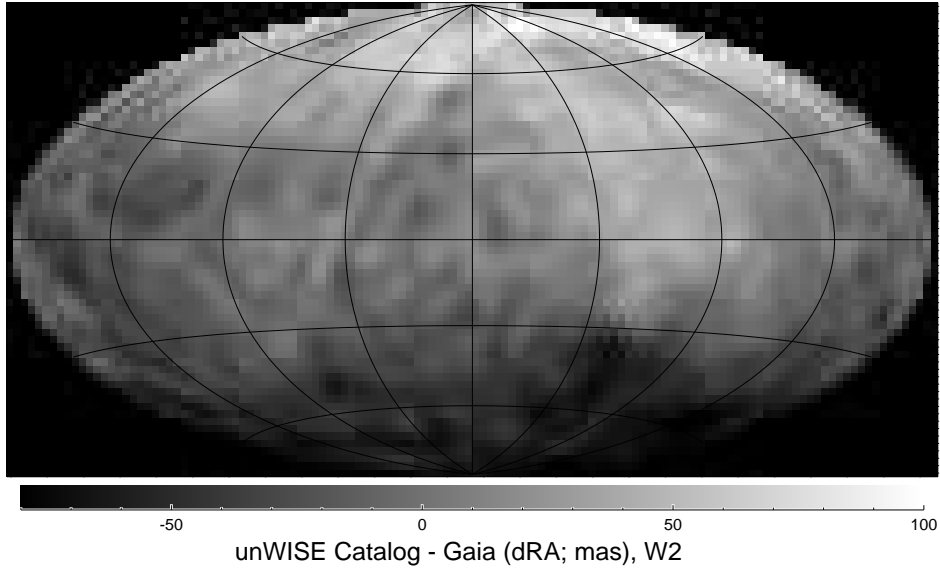}
      \caption{Full-sky map in Galactic coordinates of the five-year unWISE W2 full-depth $\Delta\alpha$ astrometric offsets relative to Gaia DR2 calibrator positions propagated to the mean unWISE epoch. The corresponding map for W1 looks similar. The large angular scale residuals shown in this figure can be eliminated by using our `recalibrated' WCS solutions. The projection is Aitoff equal-area, with ($l_{gal}$, $b_{gal}$) = (0, 0) at the center.}
       \label{fig:w2_gaia_offset_map_dra}
\end{centering}
\end{figure*}

\begin{figure*}
\begin{centering}
       \includegraphics[width=5in]{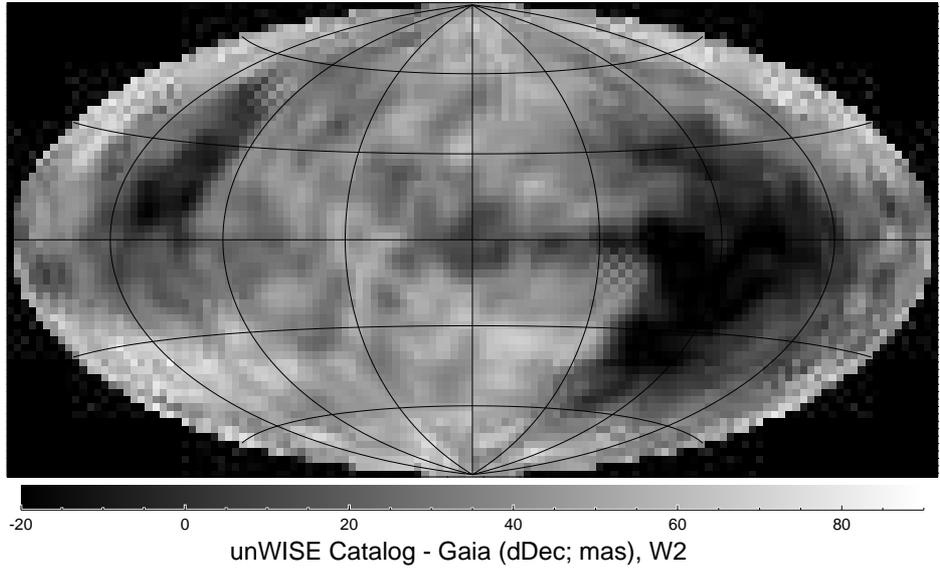}
      \caption{Full-sky map in Galactic coordinates of the five-year unWISE W2 full-depth $\Delta\delta$ astrometric offsets relative to Gaia DR2 calibrator positions propagated to the mean unWISE epoch. The corresponding map for W1 looks similar. The large angular scale residuals shown in this figure can be eliminated by using our `recalibrated' WCS solutions. The projection is Aitoff equal-area, with ($l_{gal}$, $b_{gal}$) = (0, 0) at the center.}
       \label{fig:w2_gaia_offset_map_ddec}
\end{centering}
\end{figure*}

For each full-depth unWISE coadd, we match the unWISE and Gaia sources using a radius of 2$''$, and compute offsets $\Delta x$ = median($x_{unwise} - x_{gaia}$), $\Delta y$ = median($y_{unwise} - y_{gaia}$) in terms of unWISE pixel coordinates. We also compute $\Delta \alpha$ = median(($\alpha_{unwise} - \alpha_{gaia}$)$\times$cos($\delta_{gaia}$)) and $\Delta \delta$ = median($\delta_{unwise} - \delta_{gaia}$). Figure \ref{fig:w2_gaia_offset_map_dra} (\ref{fig:w2_gaia_offset_map_ddec}) shows a full-sky map of $\Delta\alpha$ ($\Delta\delta$) in Galactic coordinates for W2, where the per-coadd values have been binned into $N_{side}$ = 16 HEALPix maps \citep{healpix} which are then smoothed with a 5$^{\circ}$ FWHM Gaussian kernel. The median number of unWISE-Gaia matches per coadd is 17,600 (13,500) in W1 (W2), varying between $\sim$6,100 (4,900) at very high Galactic latitude and a maximum of $\sim$196,000 (174,000) toward the inner Galaxy.

The large angular scale trends shown in Figures \ref{fig:w2_gaia_offset_map_dra}-\ref{fig:w2_gaia_offset_map_ddec} mainly reflect Galactic rotation and the solar apex motion. This results from the fact that pre-hibernation L1b images were calibrated to 2MASS without a proper motion correction. As a result, our maps of $\Delta\alpha$ and $\Delta\delta$ look much like Figures 1\footnote{\tiny{\url{http://wise2.ipac.caltech.edu/docs/release/allwise/expsup/figures/sec5_2biif1.png}}}  and 2\footnote{\tiny{\url{http://wise2.ipac.caltech.edu/docs/release/allwise/expsup/figures/sec5_2biif2.png}}}  of the AllWISE explanatory supplement $\S$V.2.b.ii \citep{cutri13}, albeit substantially reduced in amplitude because our full-depth coadds average together pre and post hibernation imaging, with post-reactivation L1b WCS determined in a way that does account for proper motion of the calibrator sources using UCAC4 \citep{ucac4}.

\begin{figure*}
\begin{centering}
       \includegraphics[width=5in]{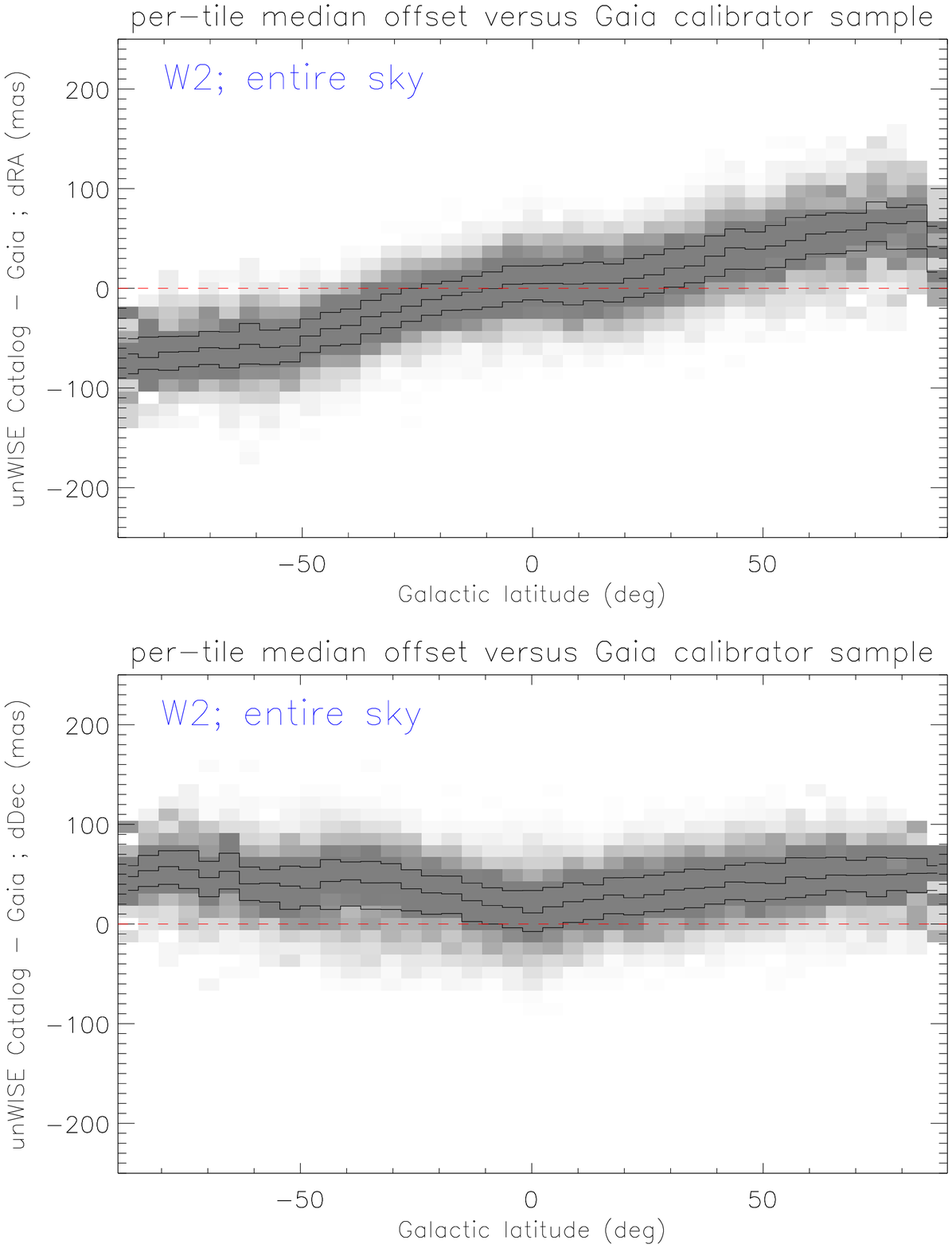}
      \caption{Trends of (unWISE $-$ Gaia) per-tile astrometric offsets as a function of Galactic latitude, for W2. Top: $\Delta\alpha$. Bottom: $\Delta\delta$. The corresponding plots for W1 look similar. The ramping of $\Delta\alpha$ (top panel) and overall $\Delta\delta$ offset (bottom panel) can be eliminated by using our `recalibrated' WCS solutions.}
       \label{fig:w2_gaia_offset_hogg}
\end{centering}
\end{figure*}

To provide a better sense for the behavior of the measured per-tile unWISE-Gaia offsets, Figure \ref{fig:w2_gaia_offset_hogg} shows plots of their trends in W2 as a function of Galactic latitude. $\Delta\alpha$ ramps between $\sim -70$ mas and $\sim +60$ mas, while $\Delta\delta$ mainly shows an overall offset of $\sim 40$ mas. Very similar trends are present in W1.

\subsubsection{Recalibrated full-depth WCS}
\label{sec:crpix}
We use the per-coadd offsets $\Delta x$ and $\Delta y$ to create `recalibrated' WCS solutions for all of the five-year full-depth unWISE coadds. For each coadd
identified by its unique (\verb|coadd_id|, band) pair, we simply update the CRPIX reference pixel coordinates by assigning CRPIX$1_{recalib}$ = CRPIX$1_{0}$ + $\Delta x$, CRPIX2$_{recalib}$ = CRPIX$2_{0}$ + $\Delta y$. CRPIX?$_{recalib}$ are the recalibrated CRPIX values, and CRPIX?$_{0}$ are the original CRPIX values, which are always exactly 1024.5. We provide the updated WCS solutions as part of our data release ($\S$\ref{sec:dr}) in an index table called \verb|fulldepth_neo4_index.fits|, which is meant to be analogous to unWISE index tables we have published previously to report recalibrated astrometry for our time-resolved coadds \citep[Table 1]{tr_neo2}. Table \ref{tab:index} provides column descriptions for the \verb|fulldepth_neo4_index.fits| index table. In addition to the recalibrated WCS solution for each coadd, this index table contains related metadata such as the mean MJD values used to propagate Gaia positions to the unWISE epoch (\verb|MJDMEAN|) and the number of unWISE-matched Gaia astrometric calibration sources per coadd (\verb|N_CALIB|).

\begin{table*}
        \centering
        \caption{Column descriptions for full-depth coadd recalibrated WCS index table.}
        \label{tab:index}
        \begin{tabular}{ll}
                \hline
                Column & Description \\
                 \hline
                COADD\_ID & \verb|coadd_id| astrometric footprint identifier \\
                RA & tile center right ascension (degrees) \\
                DEC & tile center declination (degrees) \\
                BAND & integer WISE band; either 1 or 2  \\
                LGAL & Galactic longitude corresponding to tile center (degrees) \\
                BGAL & Galactic latitude corresponding to tile center (degrees) \\
                LAMBDA & ecliptic longitude corresponding to tile center (degrees) \\
                BETA & ecliptic latitude corresponding to tile center (degrees) \\
                MJDMIN & MJD value of earliest contributing exposure \\
                MJDMAX & MJD value of latest contributing exposure \\
                MJDMEAN & mean MJD of contributing exposures \\
                DT & difference of MJDMAX and MJDMIN (days) \\
                COVMIN & minimum integer coverage in unWISE \verb|-n-u| coverage map \\
                COVMAX & maximum integer coverage in unWISE \verb|-n-u| coverage map \\
                COVMED & median integer coverage in unWISE \verb|-n-u| coverage map \\
                N\_EXP & number of exposures contributing to the coadd \\
                N\_CALIB & number of sources used for Gaia-based astrometric recalibration \\
                NAXIS & 2-elements NAXIS array for WCS \\
                CD & 2$\times$2 CD matrix for WCS \\
                CDELT & 2-element CDELT array for WCS \\
                CRPIX & 2-element CRPIX array for WCS, incorporating Gaia-based corrections \\
                CRVAL & 2-element CRVAL array for WCS \\
                CTYPE & 2-element CTYPE array for WCS \\
                LONGPOLE & LONGPOLE parameter for WCS \\
                LATPOLE & LATPOLE parameter for WCS \\
                PV2 & 2-element PV2 array for WCS \\
                \hline
        \end{tabular}
\end{table*}

\subsubsection{Validation of recalibrated full-depth WCS}
\begin{figure*}
\begin{centering}
       \includegraphics[width=5in]{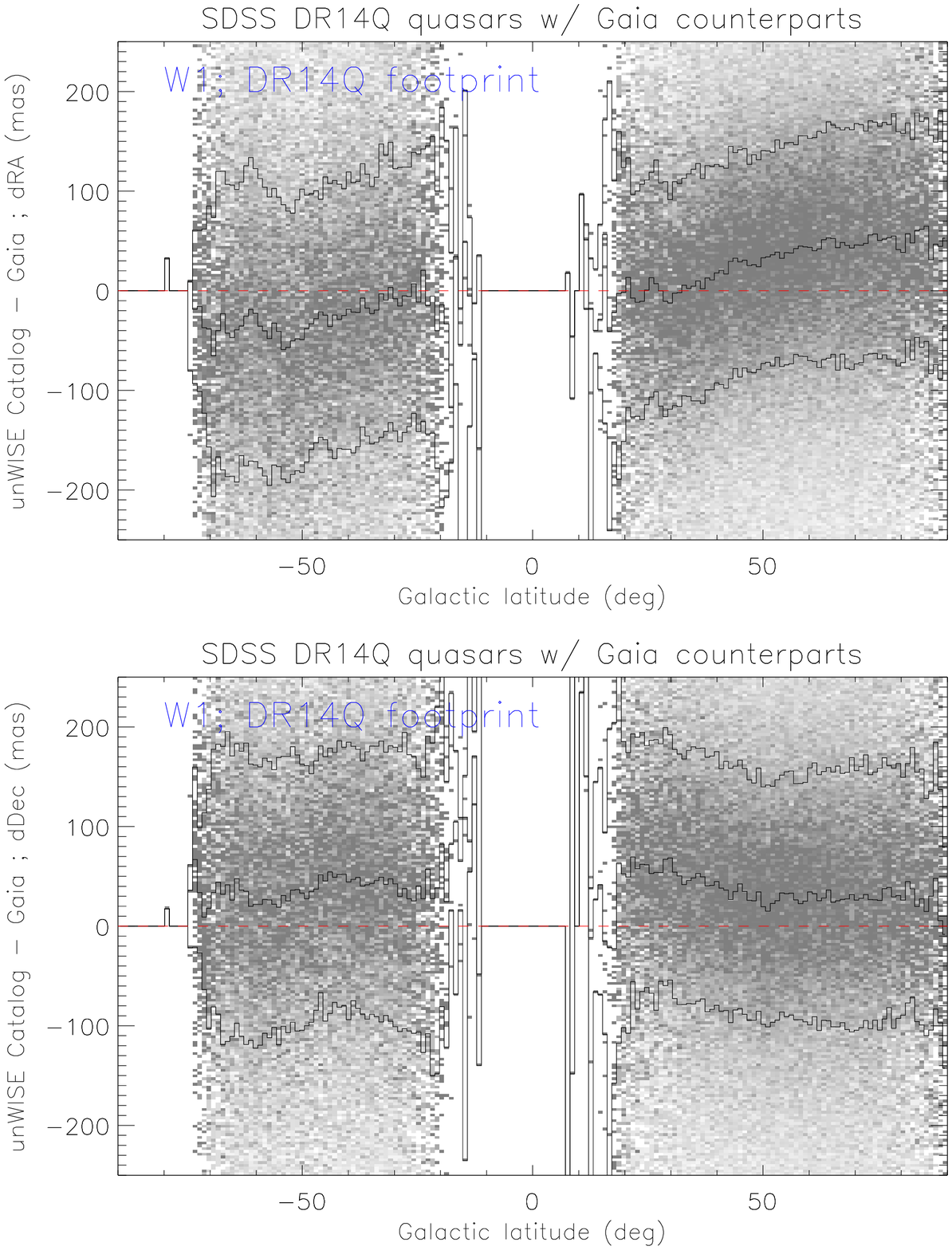}
      \caption{W1 unWISE Catalog \citep{unwise_catalog} astrometric residuals relative to Gaia as a function of Galactic latitude, for SDSS DR14Q spectroscopically confirmed quasars. Top: RA residuals. Bottom: Dec residuals. The unWISE Catalog did not incorporate the recalibrated WCS solutions of $\S$\ref{sec:crpix}. As a result, the trends shown here are similar to those of the per-coadd Gaia offsets shown in Figure \ref{fig:w2_gaia_offset_hogg}, although we should not expect exactly the same trends given that the DR14Q quasars sample a footprint which covers only $\sim$1/3 of the sky.}
       \label{fig:dr14q_quasars_w1}
\end{centering}
\end{figure*}

\begin{figure*}
\begin{centering}
       \includegraphics[width=5in]{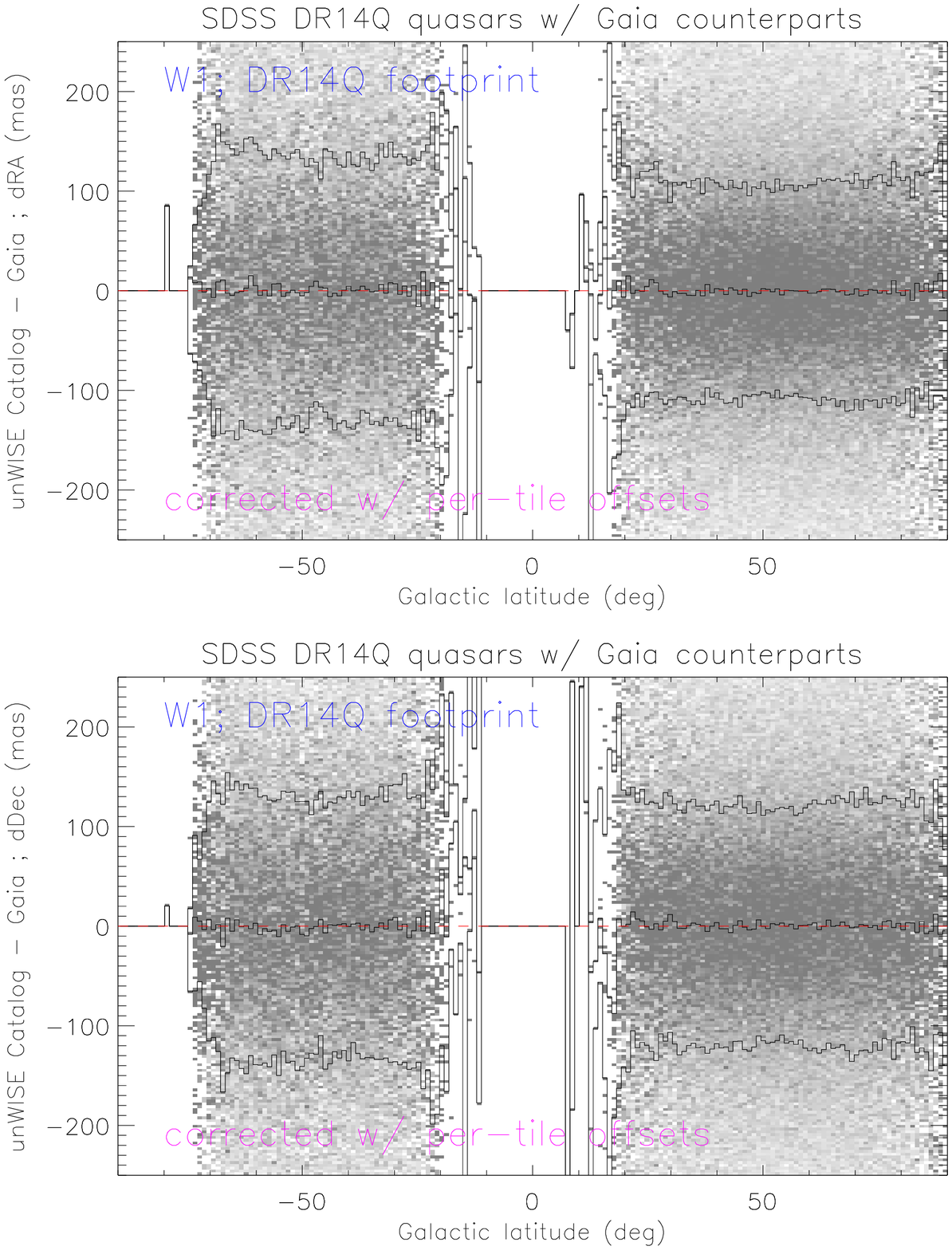}
      \caption{Residuals of recalibrated W1 unWISE Catalog positions relative to Gaia as a function of Galactic latitude, for SDSS DR14Q spectroscopically confirmed quasars. Top: RA residuals. Bottom: Dec residuals. The systematic differences between unWISE and Gaia shown in Figure \ref{fig:dr14q_quasars_w1} have been removed by converting unWISE pixel coordinates to world coordinates using our recalibrated WCS solutions.}
       \label{fig:dr14q_quasars_w1_corrected}
\end{centering}
\end{figure*}

To verify that improved astrometry results from our recalibrated WCS solutions, we analyze a sample of spectroscopically confirmed quasars drawn from the SDSS ``DR14Q'' quasar catalog \citep{dr14q}. These quasars serve as a useful validation sample because, for our purposes, they can safely be assumed stationary (negligible parallax and proper motion). We downselect to quasars that have astrometry available from Gaia DR2, retaining DR14Q quasars with a Gaia counterpart within 2$''$. $\sim$355,000 of $\sim$526,000 DR14Q quasars have such Gaia DR2 counterparts. Henceforward, for the DR14Q-Gaia sample of $\sim$355,000 quasars, we always adopt the Gaia DR2 positions as ground truth. We match the DR14Q-Gaia quasar sample against the unWISE Catalog using a 2$''$ radius, and compute per-object residuals in RA and Dec. The RA residuals include a multiplication by cos(Dec) so that they are in true angular units rather than being simple RA coordinate differences. Note that the unWISE Catalog astrometry does not take into account the WCS recalibration described in $\S$\ref{sec:crpix}. Figure \ref{fig:dr14q_quasars_w1} plots the unWISE Catalog W1 quasar astrometry residuals as a function of Galactic latitude. As anticipated, these trends are similar to those shown for the per-tile unWISE-Gaia offsets in Figure \ref{fig:w2_gaia_offset_hogg}, although we should not expect exactly the same trends given that the quasars sample a footprint which covers only $\sim$1/3 of the sky. Figure \ref{fig:dr14q_quasars_w1_corrected} shows the corresponding W1 quasar astrometry residuals upon converting the unWISE Catalog pixel coordinate centroids to (RA, Dec) with the recalibrated WCS solutions of $\S$\ref{sec:crpix}. Figure \ref{fig:dr14q_quasars_w1_corrected} illustrates that systematic deviations of the quasar positions with respect to Gaia have been removed thanks to the recalibrated WCS solutions. Analogous plots of the W2 quasar astrometry residuals before and after recalibration show very similar results.

The Gaia-DR14Q quasars are not especially bright in WISE (median Vega magnitudes of W1 $\approx$ 16.3, W2 $\approx$ 15.2), and their Gaia astrometry may contain excess scatter because these objects have host galaxies which are likely often resolved by Gaia. Therefore, to gauge the bright end scatter of the unWISE astrometry after recalibration, we assemble a sample of ``low-motion'' stars in Gaia DR2, which are meant to be brighter analogs of our Gaia-DR14Q quasar sample. Specifically, we select Gaia DR2 sources with full astrometric solutions, \verb|parallax| $< 3$ mas, \verb|parallax_error| $< 1$ mas, $|$\verb|pmra|$|$ $< 1$ mas/yr, \verb|pmra_error| $< 1$ mas/yr, $|$\verb|pmdec|$|$ $< 1$ mas/yr, and \verb|pmdec_error| $< 1$ mas/yr. We then cross-match these Gaia sources to the unWISE Catalog with a radius of 2$''$. We downselect to relatively WISE-bright sources which can still be confidently assumed unsaturated, with unWISE Catalog mags between 10 and 11.6 (Vega) in the band under consideration. We further require $|b_{gal}| > 20^{\circ}$ and unWISE Catalog \verb|fracflux| $> 0.999$. This results in Gaia-unWISE comparison samples of $\sim$17,000 (18,000) sources in W1 (W2). For these samples the median offsets before recalibration are $-$9 mas ($-$8 mas) in dRA in W1 (W2), and 29 mas (34 mas)  in dDec. After recalibration, these median offsets are $-$1 mas (0 mas) in dRA in W1 (W2), and 0 mas ($-$2 mas) in dDec. In W1, the bright end scatter as measured by the robust RMS is 58 mas (45 mas) in dRA before (after) recalibration and 52 mas (42 mas) in dDec. In W2, the bright end scatter is 52 mas (37 mas) in dRA before (after) recalibration and 47 mas (36 mas) in dDec.

In summary, considering that the W1/W2 FWHM is $\sim$6$''$, we find that the bright end scatter provided by our full-depth coadd astrometry is very good, even without employing recalibrated WCS solutions: $\sim$1/150 (1/110) of a FWHM with (without) recalibration. There are large angular scale systematics in the native full-depth unWISE coadd astrometry at the level of $\sim$150 mas peak-to-peak ($\sim$0.025 FWHM), and these systematics can be entirely eliminated by using our recalibrated WCS solutions.

\subsection{Time-resolved unWISE coadd astrometry}

Because we expect the signature scientific application of our time-resolved unWISE coadds to be motion-based discovery of faint/cold brown dwarfs not detectable by Gaia \citep[e.g.,][]{backyard_worlds}, it is critical to understand and optimize the astrometric properties of these coadds. WISE brown dwarf motion searches are often limited by the need to scrutinize $\sim$$50,000$$-$$1,000,000$ finder charts, pinpointing the small number of real moving objects among a vastly larger candidate pool dominated by false positives \citep[e.g.,][]{allwise_motion_survey, luhman_planetx, allwise2_motion_survey, schneider_neowise}. In such scenarios, both the candidate selection and vetting processes benefit to the extent that true astrophysical motion can be best distinguished from spurious apparent motion due to slight astrometric misalignments between WISE images at different epochs. Additionally, substellar object characterization can leverage high-fidelity WISE astrometry, for example in determining moving group membership to obtain age/mass constraints \citep[e.g.,][]{simp0136}.

In \cite{tr_neo2} we presented an extensive astrometric evaluation of the time-resolved unWISE coadds and showed that a bright end scatter of $\sim$50 mas per coordinate could be achieved with a coadd-level WCS recalibration following coaddition. That study was performed prior to the release of Gaia DR2, and hence we used the HSOY catalog \citep{hsoy} to obtain calibrator sources with high-quality astrometry including proper motions. For our five-year set of time-resolved coadds, we employ essentially the same astrometric recalibration procedure as in \cite{tr_neo2}, but we now replace HSOY with Gaia DR2 as our calibrator catalog. Here we provide astrometric recalibrations based on Gaia DR2 for all time-resolved unWISE coadds, not just those coadds which are built from the most recently processed year of NEOWISE-R exposures. These recalibrated WCS solutions for the time-resolved coadds are derived using SCAMP \citep{scamp}.

In selecting our sample of Gaia astrometric calibrators, we require a full 5-parameter solution so that proper motion is available for propagation of coordinates to each time-resolved coadd's epoch. The number of Gaia DR2 SCAMP calibrators per time-resolved coadd is typically quite similar to the number previously available when using HSOY. The mean (median) factor by which the number of SCAMP calibrators per coadd increases for Gaia DR2 relative to HSOY is 1.13 (1.00). The larger mean number of SCAMP calibrators with Gaia DR2 arises primarily from crowded fields in the Galactic plane, where HSOY suffers from relying in part on ground-based catalogs. At high Galactic latitude, ($|b_{gal}| > 30^{\circ}$), there are typically $\sim$2,500 calibrator sources contributing to a given coadd's SCAMP solution.

For each of the 387,915 time-resolved coadds, we gather the set of overlapping Gaia DR2 calibrators and calculate the Gaia coordinates at the relevant epoch. We then compute a first order SCAMP solution for each coadd by comparing flux-weighted centroids measured from the time-resolved unWISE coadd versus Gaia calibrator positions. Only 966 SCAMP recalibration failures occurred, representing just $\sim$0.25\% of time-resolved coadds. These rare failures generally correspond to time-resolved coadds that consist mostly of ``empty'' regions with zero integer frame coverage \citep[see Figure 7 of][]{tr_neo2}. We compared our new SCAMP recalibrations based on Gaia DR2 against those based on HSOY for tens of thousands of coadds where both are available, and the results are extremely similar. With recalibration based on Gaia DR2, our time-resolved coadds achieve an excellent bright scatter of $\sim$50 mas per coordinate, corresponding to $\sim$1/55 of the WISE pixel sidelength. The results of our time-resolved coadd WCS recalibrations are available as part of our five-year unWISE coadd data release (see $\S$\ref{sec:dr} for details).

In this work we did not make use of the Gaia calibrator parallaxes when propagating Gaia coordinates to each time-resolved coadd's epoch, so our astrometric recalibrations do not provide absolute astrometry. Our Gaia DR2 calibrator sources typically have a parallax of $\sim$1-1.5 mas, so we did not foresee any clearcut means to conclusively validate that a parallax correction had in fact been successful. In the future we will endeavor to incorporate Gaia parallaxes into our unWISE astrometric recalibration analyses.

\section{Overview of Artifact Flagging Enhancements}
\label{sec:artifact_summary}

In \cite{fulldepth_neo2} we first introduced simplistic bright star masks to the set of unWISE coadd data products. These masks exist as a set of 18,240 images which together cover the entire sky. There is one unWISE bitmask image per unWISE tile footprint, combining artifact flagging information about the W1 and W2 bands into a single file. Currently, the unWISE bitmasks do not address the W3 or W4 bands, although extending them to do so is conceivable.

The original unWISE bitmasks from \cite{fulldepth_neo2} contained only four bits. As part of the present five-year unWISE coadd data release, we have dramatically enhanced the set of features now available in our unWISE bitmasks. The total number of mask bits has increased from 4 to 31, and much more work has gone into ensuring that the masking behaves properly as a function of sky location (e.g., at low Galactic latitude and high ecliptic latitude\footnote{We will use the symbol $\beta$ to denote ecliptic latitude throughout this paper.}).

The goal of the newly upgraded unWISE bitmasks is to provide a general purpose artifact flagging capability comparable to that of the WISE team's ``CC flags'', which were generated by a module called ARTID\footnote{\scriptsize{\url{http://wise2.ipac.caltech.edu/docs/release/allsky/expsup/sec4_4g.html}}; ARTID computes flags at the catalog level and did not produce image-level renderings like those of the unWISE bitmasks.}.  The unWISE bitmasks are also constructed with particular emphasis on enabling rare object searches. Examples of such applications for which high-quality WISE artifact flagging plays an essential role are searches for exceptionally cold brown dwarfs and very high redshift quasars \citep[e.g.,][]{banados_quasar}. Most but not all of the unWISE mask bits are associated with artifacts from bright stars (see Table \ref{tab:bitmask}). The unWISE bitmasks are integer-valued and populated with sums of powers of 2, such that a particular bit can be isolated by taking the bitwise AND between the mask image and the desired power of 2. The mask bit definitions in Table \ref{tab:bitmask} are also documented in the FITS header of each unWISE bitmask image. Appendix \ref{app:bitmask} fully explains the meanings of all mask bits, how the different bits are related to one another, and the detailed procedure for creating each unWISE mask bit. Our enhanced unWISE bitmasks are already in use by the unWISE Catalog \citep{unwise_catalog}, the DESI pre-imaging surveys \citep[DR8 and onward;][]{dey_overview}, and the CatWISE motion catalog \citep{catwise_preliminary}.

\begin{table}
        \centering
        \caption{Definitions of unWISE artifact flagging mask bits.}
        \label{tab:bitmask}
        \begin{tabular}{llc}
                \hline
                Bit & Description \\
                 \hline
                0 & W1 bright star, southward scan \\
                1 & W1 bright star, northward scan \\
                2 & W2 bright star, southward scan \\
                3 & W2 bright star, northward scan \\
                4 & W1 bright star saturation \\
                5 & W2 bright star saturation \\
                6 & center of pixel not primary \\
                7 & W1 bright star, centroid off edge \\
                8 & W2 bright star, centroid off edge \\
                9 & resolved galaxy \\
                10 & big object (LMC, SMC, M31) \\
                11 & W2 optical ghost, southward scan \\
                12 & W2 optical ghost, northward scan \\
                13 & W1 first latent, southward scan \\
                14 & W1 first latent, northward scan \\
                15 & W2 first latent, southward scan \\
                16 & W2 first latent, northward scan \\
                17 & W1 second latent, southward scan \\
                18 & W1 second latent, northward scan \\
                19 & W2 second latent, southward scan \\
                20 & W2 second latent, northward scan \\
                21 & may contain W1 bright star centroid \\
                22 & may contain W2 bright star centroid \\
                23 & AllWISE-like W1 circular halo \\
                24 & AllWISE-like W2 circular halo \\
                25 & W1 optical ghost, southward scan \\
                26 & W1 optical ghost, northward scan \\
                27 & PSF-based W1 diffraction spike \\
                28 & PSF-based W2 diffraction spike \\
                29 & geometric W1 diffraction spike \\
                30 & geometric W2 diffraction spike \\
                \hline
        \end{tabular}
\end{table}

The unWISE bitmasks are optimized for catalogs that use detailed ``pixelized'' WISE PSF modeling and have a flexible WISE sky background model. To the extent that these capabilities are not implemented in a given WISE catalog's construction, the unWISE bitmasks may underflag features such as diffraction spikes and background level variations due to scattered light. Some artifacts are assigned separate unWISE mask bits for northward versus southward scan directions; this may be useful for time domain applications. The fraction of area flagged by any unWISE mask bit in either W1 or W2 ramps up from $\sim$1\% at Galactic latitude of 90$^{\circ}$ to $\sim$5\% at the lowest Galactic latitude edge of the extragalactic sky (as defined by the DESI cosmology survey footprint's boundary, $|b_{gal}| \sim 18^{\circ}$).

\section{Validation of the Five-year unWISE Coadds}
\label{sec:validation}

\begin{figure*}
\begin{center}
\includegraphics[width=7.0in]{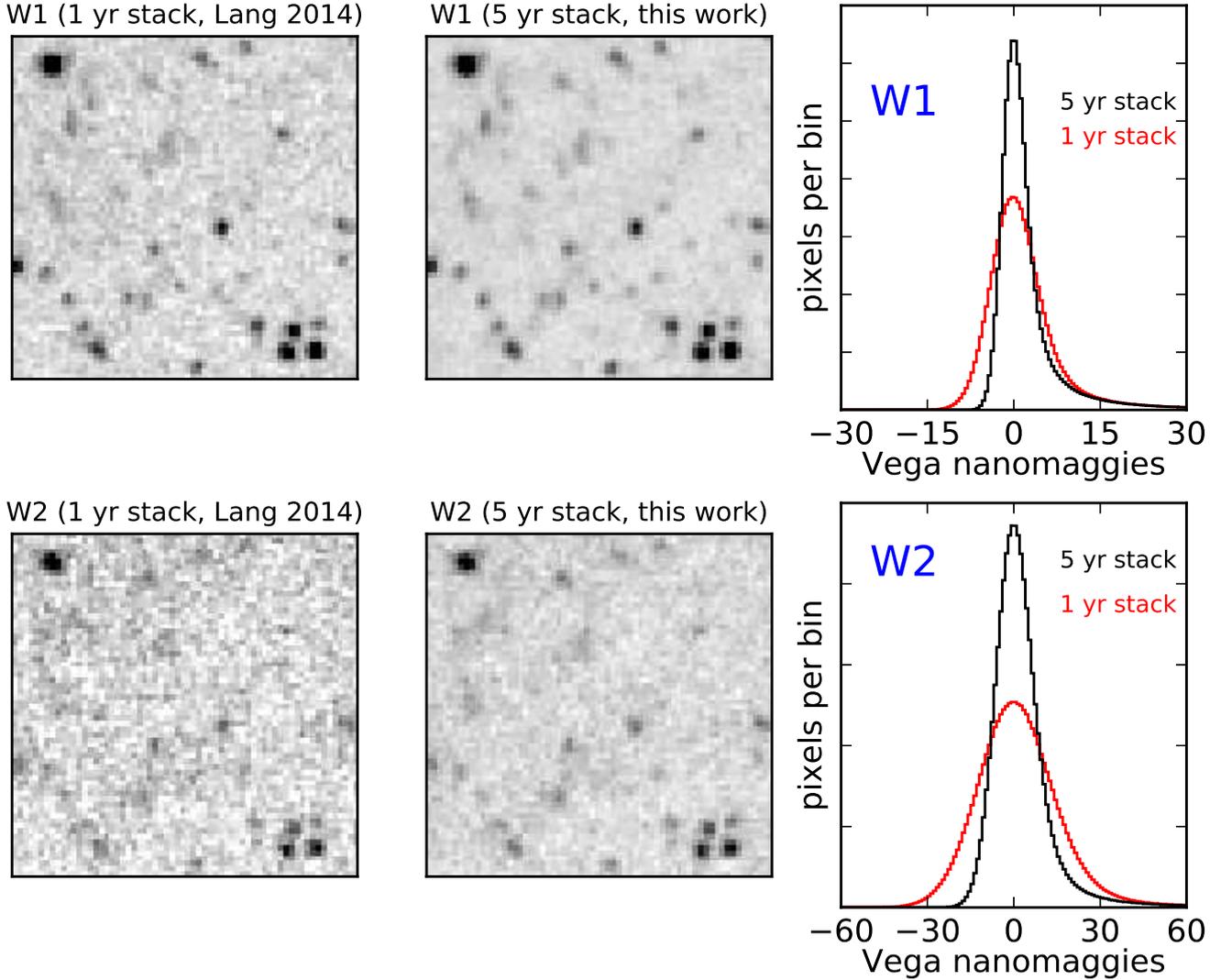}
\caption{Depiction of reduced pixel noise in our five-year full-depth coadds relative to those incorporating only pre-hibernation W1/W2 exposures. Top row: W1. Bottom row: W2. Left column: unWISE coadds built from pre-hibernation W1/W2 imaging \citep{lang14}. Middle column: this work's corresponding five-year full-depth coadds. Each cutout is 2.9$'$ on a side, centered at $(\alpha, \delta)$ = (149.62$^{\circ}$, 1.45$^{\circ}$). This is a typical patch of sky at low ecliptic latitude and high Galactic latitude. Right column: histograms of pixel values for the entire $\sim$2.5 deg$^2$ unWISE tile footprint from which the cutouts are drawn, suggesting that a $\sim$2$\times$ decrease in random pixel noise is attained by making use of $5\times$ more input single-exposure imaging.}
\label{fig:pixel_hist}
\end{center}
\end{figure*}

The improved depth of our five-year full-depth unWISE coadds can be readily discerned by comparing the pixel noise present in these new coadds versus those which only use pre-hibernation WISE data, such as the AllWISE Atlas stacks and the original \cite{lang14} unWISE coadds. Figure \ref{fig:pixel_hist} illustrates this decreased pixel noise, and hence increased depth, by displaying coadd cutouts of a small sky patch in the COSMOS region, a typical area of extragalactic sky at low ecliptic latitude. Comparison of pixel value  histograms for the 1 yr versus 5 yr unWISE stacks on the full COSMOS tile footprint (\verb|coadd_id| = 1497p015) also shows that our new stacks have $\sim$2$\times$ tighter distributions, consistent with being $\sim$2$\times$ deeper in terms of their 5$\sigma$ flux limit. Because such histograms are substantially skewed by the presence of compact sources, it is not ideal to use these in quoting a precise estimate for the depth improvement.

The unWISE Catalog \citep{unwise_catalog} is based on the same five-year unWISE coadds presented in this work, and so the excellent catalog-level photometric performance already demonstrated by \cite{unwise_catalog} can be considered a thorough validation of our new coadds. The \cite{unwise_catalog} Figure 4 comparison to Spitzer truth catalogs in the COSMOS region \citep{scosmos} demonstrates that our new coadds support source detection to 0.76 (0.67) mag deeper than does  pre-hibernation WISE data alone in W1 (W2) --- indeed a $\sim$2$\times$ depth enhancement. \cite{unwise_catalog} also shows that for unsaturated sources, the photometry based on our five-year coadds is linear with that of AllWISE to within $\sim$3\% over a $\sim$9 mag range in source brightness. The unWISE Catalog displays overall photomeric offsets relative to AllWISE of 4 mmag (32 mmag) in W1 (W2), but these could be due to PSF model normalization choices rather than the multiplicative scaling of the five-year unWISE coadds themselves.

We have already presented a detailed validation of the astrometric performance of our five-year unWISE coadds ($\S$\ref{sec:astrometry}), concluding that with recalibration to Gaia DR2, the full-depth and time-resolved coadds both display a bright-end scatter of just $\sim$50 mas per coordinate, equivalent to $\sim$1/55 of a WISE pixel ($\sim$1/120 of the W1/W2 FWHM).

\section{Data Release}
\label{sec:dr}

The five-year unWISE coadd data release is publicly available online\footnote{\url{http://neo4-coadds.github.io}}. The data release consists of a set of FITS files organized into a hierarchy of directories. Within the top-level data release directory, a subdirectory called \verb|fulldepth/| contains the all-sky set of five-year full-depth unWISE coadds. The full-depth coadd outputs for each \verb|coadd_id| are contained within a directory named after that \verb|coadd_id|, with an intermediate level of directories named after the first three digits of the \verb|coadd_id|. For example, both bands of \verb|coadd_id| = 1497p015 full-depth products are in \verb|fulldepth/149/1497p015|. The unWISE bitmask files discussed in $\S$\ref{sec:artifact_summary} and Appendix \ref{app:bitmask} are those with names ending in \verb|-msk.fits.gz|. The five-year full-depth coadds can also be accessed via the \url{http://unwise.me} cutout interface by selecting the ``NeoWISE-R 4'' option.

The time-resolved coadds can be found in a series of top-level subdirectories \verb|e000|, \verb|e001|, ..., \verb|e184|, where the trailing three digits encode the unWISE epoch number\footnote{See $\S$3.2.2 of \cite{tr_neo2} for a precise definition of the unWISE time-resolved coadd epoch numbering scheme.}. Analogous to the full-depth directory structure, epoch 0 of \verb|coadd_id| = 1497p015 is in a directory called \verb|e000/149/1497p015|.

The recalibrated WCS solutions described in $\S$\ref{sec:astrometry} are provided in FITS binary tables named \verb|tr_neo4_index.fits| and \verb|fulldepth/fulldepth_neo4_index.fits|. The data model of the former is provided in Table 1 of \cite{tr_neo2}, and the latter's data model is specified in Table \ref{tab:index}.

The total data volume of the five-year full-depth (time-resolved) unWISE coadds is 2.9 TB (25.9 TB). The WiseView visualization tool \citep{wiseview} provides a seamless browser-based interface for exploring cutouts and blinks of both the full-depth and time-resolved unWISE coadds without needing to download large volumes of FITS images.

\section{Conclusion}
\label{sec:conclusion}

We have presented a new set of full-sky coadded maps based on the first five years of W1 and W2 imaging provided by the WISE and NEOWISE missions. Our new full-depth coadds now constitute the deepest ever all-sky maps at 3$-$5$\mu$m, enabling detection of sources $\sim$2$\times$ fainter than AllWISE at 5$\sigma$ significance. Our full-sky set of time-resolved unWISE coadds provides a 7.5 year time baseline for measuring variability and motions of sources well below the single-exposure detection limit, a 15$\times$ enhancement relative to the 0.5 year AllWISE time baseline. Relative to our prior (four-year) set of unWISE coadds, the full-depth stacks presented here provide a 25\% increase in total exposure time and up to an 18\% increase in the signal-to-noise of proper motion measurements. In addition to folding in an extra fifth year of W1/W2 imaging, we have also presented improvements to the coadd-level astrometric calibration and unWISE artifact flagging capabilities. The new image-level data products described in this work are already being used by several ambitious wide-area cataloging efforts: CatWISE, the unWISE Catalog, and the DESI pre-imaging surveys.

More archival analysis will be required to maximize the complete WISE/NEOWISE data set's value for Galactic and extragalactic astrophysics. A sixth year of W1/W2 imaging recently became publicly available in 2019 April, and these new data ought to be incorporated into future versions of the unWISE coadds and derivative catalogs.

\section*{Acknowledgements}

This work has been supported by grant NNH17AE75I from the NASA Astrophysics Data Analysis Program. AMM acknowledges support from Hubble Fellowship HST-HF2-51415.001-A. We thank John Moustakas for sharing early versions of his LSLGA catalog with us.

This research makes use of data products from the Wide-field Infrared Survey Explorer, which is a joint project of the University of California, Los Angeles, and the Jet Propulsion Laboratory/California Institute of Technology, funded by the National Aeronautics and Space Administration. This research also makes use of data products from NEOWISE, which is a project of the Jet Propulsion Laboratory/California Institute of Technology, funded by the Planetary Science Division of the National Aeronautics and Space Administration. This research has made use of the NASA/ IPAC Infrared Science Archive, which is operated by the Jet Propulsion Laboratory, California Institute of Technology, under contract with the National Aeronautics and Space Administration.

The National Energy Research Scientific Computing Center, which is supported by the Office of Science of the U.S. Department of Energy under Contract No. DE-AC02-05CH11231, provided staff, computational resources, and data storage for this project.

\bibliographystyle{mnras}
\bibliography{neo4}

\begin{thebibliography}{}
\makeatletter
\relax
\def\mn@urlcharsother{\let\do\@makeother \do\$\do\&\do\#\do\^\do\_\do\%\do\~}
\def\mn@doi{\begingroup\mn@urlcharsother \@ifnextchar [ {\mn@doi@}
  {\mn@doi@[]}}
\def\mn@doi@[#1]#2{\def\@tempa{#1}\ifx\@tempa\@empty \href
  {http://dx.doi.org/#2} {doi:#2}\else \href {http://dx.doi.org/#2} {#1}\fi
  \endgroup}
\def\mn@eprint#1#2{\mn@eprint@#1:#2::\@nil}
\def\mn@eprint@arXiv#1{\href {http://arxiv.org/abs/#1} {{\tt arXiv:#1}}}
\def\mn@eprint@dblp#1{\href {http://dblp.uni-trier.de/rec/bibtex/#1.xml}
  {dblp:#1}}
\def\mn@eprint@#1:#2:#3:#4\@nil{\def\@tempa {#1}\def\@tempb {#2}\def\@tempc
  {#3}\ifx \@tempc \@empty \let \@tempc \@tempb \let \@tempb \@tempa \fi \ifx
  \@tempb \@empty \def\@tempb {arXiv}\fi \@ifundefined
  {mn@eprint@\@tempb}{\@tempb:\@tempc}{\expandafter \expandafter \csname
  mn@eprint@\@tempb\endcsname \expandafter{\@tempc}}}

\bibitem[\protect\citeauthoryear{{Altmann}, {Roeser}, {Demleitner}, {Bastian}
  \& {Schilbach}}{{Altmann} et~al.}{2017}]{hsoy}
{Altmann} M.,  {Roeser} S.,  {Demleitner} M.,  {Bastian} U.,   {Schilbach} E.,
  2017, \mn@doi [\aap] {10.1051/0004-6361/201730393}, \href
  {https://ui.adsabs.harvard.edu/abs/2017A&A...600L...4A} {600, L4}

\bibitem[\protect\citeauthoryear{{Ba{\~n}ados} et~al.,}{{Ba{\~n}ados}
  et~al.}{2018}]{banados_quasar}
{Ba{\~n}ados} E.,  et~al., 2018, \mn@doi [\nat] {10.1038/nature25180}, \href
  {https://ui.adsabs.harvard.edu/abs/2018Natur.553..473B} {553, 473}

\bibitem[\protect\citeauthoryear{{Bertin}}{{Bertin}}{2006}]{scamp}
{Bertin} E.,  2006, in {Gabriel} C.,  {Arviset} C.,  {Ponz} D.,   {Enrique} S.,
   eds,  Astronomical Society of the Pacific Conference Series Vol. 351,
  Astronomical Data Analysis Software and Systems XV. p.~112

\bibitem[\protect\citeauthoryear{{Caselden}, {Westin}, {Meisner}, {Kuchner}  \&
  {Colin}}{{Caselden} et~al.}{2018}]{wiseview}
{Caselden} D.,  {Westin} III P.,  {Meisner} A.,  {Kuchner} M.,   {Colin} G.,
  2018, {WiseView: Visualizing motion and variability of faint WISE sources},
  Astrophysics Source Code Library (\mn@eprint {ascl} {1806.004})

\bibitem[\protect\citeauthoryear{{Cutri} et~al.,}{{Cutri}
  et~al.}{2013}]{cutri13}
{Cutri} R.~M.,  et~al., 2013, Technical report, {Explanatory Supplement to the
  AllWISE Data Release Products}

\bibitem[\protect\citeauthoryear{{Cutri} et~al.,}{{Cutri}
  et~al.}{2015}]{cutri15}
{Cutri} R.~M.,  et~al., 2015, Technical report, {Explanatory Supplement to the
  NEOWISE Data Release Products}

\bibitem[\protect\citeauthoryear{{DESI Collaboration}, {Aghamousa}, {Aguilar},
  {Ahlen}, {Allen}  \& {Allende}}{{DESI Collaboration}
  et~al.}{2016a}]{desi_part1}
{DESI Collaboration} {Aghamousa} A.,  {Aguilar} J.,  {Ahlen} S.and~{Alam} S.,
  {Allen} L.,   {Allende} C.,  2016a, preprint, \href
  {http://adsabs.harvard.edu/abs/2016arXiv161100036D} {} (\mn@eprint {arXiv}
  {1611.00036})

\bibitem[\protect\citeauthoryear{{DESI Collaboration}, {Aghamousa}, {Aguilar},
  {Ahlen}, {Alam}, {Allen}  \& {Allende}}{{DESI Collaboration}
  et~al.}{2016b}]{desi_part2}
{DESI Collaboration} {Aghamousa} A.,  {Aguilar} J.,  {Ahlen} S.,  {Alam} S.,
  {Allen} L.,   {Allende} C.,  2016b, preprint, \href
  {http://adsabs.harvard.edu/abs/2016arXiv161100037D} {} (\mn@eprint {arXiv}
  {1611.00037})

\bibitem[\protect\citeauthoryear{{Dey} et~al.,}{{Dey} et~al.}{2016}]{mosaic3}
{Dey} A.,  et~al., 2016, in Ground-based and Airborne Instrumentation for
  Astronomy VI. p. 99082C, \mn@doi{10.1117/12.2231488}

\bibitem[\protect\citeauthoryear{{Dey} et~al.,}{{Dey}
  et~al.}{2019}]{dey_overview}
{Dey} A.,  et~al., 2019, \mn@doi [\aj] {10.3847/1538-3881/ab089d}, \href
  {https://ui.adsabs.harvard.edu/abs/2019AJ....157..168D} {157, 168}

\bibitem[\protect\citeauthoryear{{Dor{\'e}} et~al.,}{{Dor{\'e}}
  et~al.}{2018}]{spherex}
{Dor{\'e}} O.,  et~al., 2018, arXiv e-prints, \href
  {https://ui.adsabs.harvard.edu/abs/2018arXiv180505489D} {p. arXiv:1805.05489}

\bibitem[\protect\citeauthoryear{{Eisenhardt} et~al.,}{{Eisenhardt}
  et~al.}{2019}]{catwise_preliminary}
{Eisenhardt} P. R.~M.,  et~al., 2019, arXiv e-prints, \href
  {https://ui.adsabs.harvard.edu/abs/2019arXiv190808902E} {p. arXiv:1908.08902}

\bibitem[\protect\citeauthoryear{{Flaugher} et~al.,}{{Flaugher}
  et~al.}{2015}]{decam}
{Flaugher} B.,  et~al., 2015, \mn@doi [\aj] {10.1088/0004-6256/150/5/150},
  \href {https://ui.adsabs.harvard.edu/abs/2015AJ....150..150F} {150, 150}

\bibitem[\protect\citeauthoryear{{Gagn{\'e}} et~al.,}{{Gagn{\'e}}
  et~al.}{2017}]{simp0136}
{Gagn{\'e}} J.,  et~al., 2017, \mn@doi [\apj] {10.3847/2041-8213/aa70e2}, \href
  {https://ui.adsabs.harvard.edu/abs/2017ApJ...841L...1G} {841, L1}

\bibitem[\protect\citeauthoryear{{Gaia Collaboration} et~al.,}{{Gaia
  Collaboration} et~al.}{2018}]{gaia_dr2}
{Gaia Collaboration} et~al., 2018, \mn@doi [\aap]
  {10.1051/0004-6361/201833051}, \href
  {https://ui.adsabs.harvard.edu/abs/2018A&A...616A...1G} {616, A1}

\bibitem[\protect\citeauthoryear{{Gardner} et~al.,}{{Gardner}
  et~al.}{2006}]{jwst}
{Gardner} J.~P.,  et~al., 2006, \mn@doi [\ssr] {10.1007/s11214-006-8315-7},
  \href {https://ui.adsabs.harvard.edu/abs/2006SSRv..123..485G} {123, 485}

\bibitem[\protect\citeauthoryear{{G{\'o}rski}, {Hivon}, {Banday}, {Wandelt},
  {Hansen}, {Reinecke}  \& {Bartelmann}}{{G{\'o}rski} et~al.}{2005}]{healpix}
{G{\'o}rski} K.~M.,  {Hivon} E.,  {Banday} A.~J.,  {Wandelt} B.~D.,  {Hansen}
  F.~K.,  {Reinecke} M.,   {Bartelmann} M.,  2005, \mn@doi [\apj]
  {10.1086/427976}, \href
  {https://ui.adsabs.harvard.edu/abs/2005ApJ...622..759G} {622, 759}

\bibitem[\protect\citeauthoryear{{Kirkpatrick} et~al.,}{{Kirkpatrick}
  et~al.}{2011}]{kirkpatrick11}
{Kirkpatrick} J.~D.,  et~al., 2011, \mn@doi [\apjs]
  {10.1088/0067-0049/197/2/19}, \href
  {https://ui.adsabs.harvard.edu/abs/2011ApJS..197...19K} {197, 19}

\bibitem[\protect\citeauthoryear{{Kirkpatrick} et~al.,}{{Kirkpatrick}
  et~al.}{2014}]{allwise_motion_survey}
{Kirkpatrick} J.~D.,  et~al., 2014, \mn@doi [\apj]
  {10.1088/0004-637X/783/2/122}, \href
  {https://ui.adsabs.harvard.edu/abs/2014ApJ...783..122K} {783, 122}

\bibitem[\protect\citeauthoryear{{Kirkpatrick} et~al.,}{{Kirkpatrick}
  et~al.}{2016}]{allwise2_motion_survey}
{Kirkpatrick} J.~D.,  et~al., 2016, \mn@doi [\apjs]
  {10.3847/0067-0049/224/2/36}, \href
  {https://ui.adsabs.harvard.edu/abs/2016ApJS..224...36K} {224, 36}

\bibitem[\protect\citeauthoryear{{Kuchner} et~al.,}{{Kuchner}
  et~al.}{2017}]{backyard_worlds}
{Kuchner} M.~J.,  et~al., 2017, \mn@doi [\apjl] {10.3847/2041-8213/aa7200},
  \href {http://adsabs.harvard.edu/abs/2017ApJ...841L..19K} {841, L19}

\bibitem[\protect\citeauthoryear{{Lang}}{{Lang}}{2014}]{lang14}
{Lang} D.,  2014, \mn@doi [\aj] {10.1088/0004-6256/147/5/108}, \href
  {http://adsabs.harvard.edu/abs/2014AJ....147..108L} {147, 108}

\bibitem[\protect\citeauthoryear{{Lang}, {Hogg}  \& {Schlegel}}{{Lang}
  et~al.}{2016}]{lang14b}
{Lang} D.,  {Hogg} D.~W.,   {Schlegel} D.,  2016, \mn@doi [\aj]
  {10.3847/0004-6256/151/2/36}, \href
  {http://adsabs.harvard.edu/abs/2016AJ....151...36L} {151, 36}

\bibitem[\protect\citeauthoryear{{Levi}, {Bebek}, {Beers}, {Blum}, {Cahn}  \&
  {Eisenstein}}{{Levi} et~al.}{2013}]{desi}
{Levi} M.,  {Bebek} C.,  {Beers} T.,  {Blum} R.,  {Cahn} R.,   {Eisenstein} D.,
   2013, preprint, \href {http://adsabs.harvard.edu/abs/2013arXiv1308.0847L} {}
  (\mn@eprint {arXiv} {1308.0847})

\bibitem[\protect\citeauthoryear{{Luhman}}{{Luhman}}{2014}]{luhman_planetx}
{Luhman} K.~L.,  2014, \mn@doi [\apj] {10.1088/0004-637X/781/1/4}, \href
  {https://ui.adsabs.harvard.edu/abs/2014ApJ...781....4L} {781, 4}

\bibitem[\protect\citeauthoryear{{Mainzer} et~al.,}{{Mainzer}
  et~al.}{2011}]{neowise}
{Mainzer} A.,  et~al., 2011, \mn@doi [\apj] {10.1088/0004-637X/731/1/53}, \href
  {http://adsabs.harvard.edu/abs/2011ApJ...731...53M} {731, 53}

\bibitem[\protect\citeauthoryear{{Mainzer} et~al.,}{{Mainzer}
  et~al.}{2014}]{neowiser}
{Mainzer} A.,  et~al., 2014, \mn@doi [\apj] {10.1088/0004-637X/792/1/30}, \href
  {http://adsabs.harvard.edu/abs/2014ApJ...792...30M} {792, 30}

\bibitem[\protect\citeauthoryear{{Mainzer} et~al.,}{{Mainzer}
  et~al.}{2015}]{neocam}
{Mainzer} A.,  et~al., 2015, \mn@doi [\aj] {10.1088/0004-6256/149/5/172}, \href
  {https://ui.adsabs.harvard.edu/abs/2015AJ....149..172M} {149, 172}

\bibitem[\protect\citeauthoryear{{Makarov}, {Prugniel}, {Terekhova}, {Courtois}
   \& {Vauglin}}{{Makarov} et~al.}{2014}]{hyperleda}
{Makarov} D.,  {Prugniel} P.,  {Terekhova} N.,  {Courtois} H.,   {Vauglin} I.,
  2014, \mn@doi [\aap] {10.1051/0004-6361/201423496}, \href
  {http://adsabs.harvard.edu/abs/2014A%26A...570A..13M} {570, A13}

\bibitem[\protect\citeauthoryear{{Marocco} et~al.,}{{Marocco}
  et~al.}{2019}]{j1935}
{Marocco} F.,  et~al., 2019, \mn@doi [\apj] {10.3847/1538-4357/ab2bf0}, \href
  {https://ui.adsabs.harvard.edu/abs/2019ApJ...881...17M} {881, 17}

\bibitem[\protect\citeauthoryear{{Meisner} \& {Finkbeiner}}{{Meisner} \&
  {Finkbeiner}}{2014}]{meisner14}
{Meisner} A.,  {Finkbeiner} D.~P.,  2014, \mn@doi [\apj]
  {10.1088/0004-637X/781/1/5}, \href
  {http://adsabs.harvard.edu/abs/2014ApJ...781....5M} {781, 5}

\bibitem[\protect\citeauthoryear{{Meisner}, {Lang}  \& {Schlegel}}{{Meisner}
  et~al.}{2017a}]{fulldepth_neo1}
{Meisner} A.~M.,  {Lang} D.,   {Schlegel} D.~J.,  2017a, \mn@doi [\aj]
  {10.3847/1538-3881/153/1/38}, \href
  {http://adsabs.harvard.edu/abs/2017AJ....153...38M} {153, 38}

\bibitem[\protect\citeauthoryear{{Meisner}, {Lang}  \& {Schlegel}}{{Meisner}
  et~al.}{2017b}]{fulldepth_neo2}
{Meisner} A.,  {Lang} D.,   {Schlegel} D.,  2017b, \mn@doi [\aj]
  {10.3847/1538-3881/aa894e}, \href
  {http://adsabs.harvard.edu/abs/2017AJ....154..161M} {154, 161}

\bibitem[\protect\citeauthoryear{{Meisner}, {Lang}  \& {Schlegel}}{{Meisner}
  et~al.}{2018a}]{fulldepth_neo3}
{Meisner} A.~M.,  {Lang} D.,   {Schlegel} D.~J.,  2018a, \mn@doi [Research
  Notes of the American Astronomical Society] {10.3847/2515-5172/aaa4bc}, \href
  {http://adsabs.harvard.edu/abs/2018RNAAS...2a...1M} {2, 1}

\bibitem[\protect\citeauthoryear{{Meisner}, {Lang}  \& {Schlegel}}{{Meisner}
  et~al.}{2018b}]{tr_neo3}
{Meisner} A.~M.,  {Lang} D.~A.,   {Schlegel} D.~J.,  2018b, \mn@doi [Research
  Notes of the American Astronomical Society] {10.3847/2515-5172/aaecd5}, \href
  {http://adsabs.harvard.edu/abs/2018RNAAS...2d.202M} {2, 202}

\bibitem[\protect\citeauthoryear{{Meisner}, {Lang}  \& {Schlegel}}{{Meisner}
  et~al.}{2018c}]{tr_neo2}
{Meisner} A.~M.,  {Lang} D.,   {Schlegel} D.~J.,  2018c, \mn@doi [\aj]
  {10.3847/1538-3881/aacbcd}, \href
  {http://adsabs.harvard.edu/abs/2018AJ....156...69M} {156, 69}

\bibitem[\protect\citeauthoryear{{P{\^a}ris} et~al.,}{{P{\^a}ris}
  et~al.}{2018}]{dr14q}
{P{\^a}ris} I.,  et~al., 2018, \mn@doi [\aap] {10.1051/0004-6361/201732445},
  \href {https://ui.adsabs.harvard.edu/abs/2018A&A...613A..51P} {613, A51}

\bibitem[\protect\citeauthoryear{{Racca} et~al.,}{{Racca}
  et~al.}{2016}]{euclid}
{Racca} G.~D.,  et~al., 2016, in Space Telescopes and Instrumentation 2016:
  Optical, Infrared, and Millimeter Wave. p. 99040O (\mn@eprint {arXiv}
  {1610.05508}), \mn@doi{10.1117/12.2230762}

\bibitem[\protect\citeauthoryear{{Ross} et~al.,}{{Ross}
  et~al.}{2018}]{ross2018}
{Ross} N.~P.,  et~al., 2018, \mn@doi [\mnras] {10.1093/mnras/sty2002}, \href
  {http://adsabs.harvard.edu/abs/2018MNRAS.480.4468R} {480, 4468}

\bibitem[\protect\citeauthoryear{{Sanders} et~al.,}{{Sanders}
  et~al.}{2007}]{scosmos}
{Sanders} D.~B.,  et~al., 2007, \mn@doi [\apjs] {10.1086/517885}, \href
  {http://adsabs.harvard.edu/abs/2007ApJS..172...86S} {172, 86}

\bibitem[\protect\citeauthoryear{{Schlafly} et~al.,}{{Schlafly}
  et~al.}{2018}]{decaps}
{Schlafly} E.~F.,  et~al., 2018, \mn@doi [\apjs] {10.3847/1538-4365/aaa3e2},
  \href {https://ui.adsabs.harvard.edu/abs/2018ApJS..234...39S} {234, 39}

\bibitem[\protect\citeauthoryear{{Schlafly}, {Meisner}  \& {Green}}{{Schlafly}
  et~al.}{2019}]{unwise_catalog}
{Schlafly} E.~F.,  {Meisner} A.~M.,   {Green} G.~M.,  2019, \mn@doi [\apjs]
  {10.3847/1538-4365/aafbea}, \href
  {https://ui.adsabs.harvard.edu/abs/2019ApJS..240...30S} {240, 30}

\bibitem[\protect\citeauthoryear{{Schlegel}, {Finkbeiner}  \&
  {Davis}}{{Schlegel} et~al.}{1998}]{sfd98}
{Schlegel} D.~J.,  {Finkbeiner} D.~P.,   {Davis} M.,  1998, \mn@doi [\apj]
  {10.1086/305772}, \href {http://adsabs.harvard.edu/abs/1998ApJ...500..525S}
  {500, 525}

\bibitem[\protect\citeauthoryear{{Schneider}, {Greco}, {Cushing},
  {Kirkpatrick}, {Mainzer}, {Gelino}, {Fajardo-Acosta}  \& {Bauer}}{{Schneider}
  et~al.}{2016}]{schneider_neowise}
{Schneider} A.~C.,  {Greco} J.,  {Cushing} M.~C.,  {Kirkpatrick} J.~D.,
  {Mainzer} A.,  {Gelino} C.~R.,  {Fajardo-Acosta} S.~B.,   {Bauer} J.,  2016,
  \mn@doi [\apj] {10.3847/0004-637X/817/2/112}, \href
  {https://ui.adsabs.harvard.edu/abs/2016ApJ...817..112S} {817, 112}

\bibitem[\protect\citeauthoryear{{Spergel} et~al.,}{{Spergel}
  et~al.}{2015}]{wfirst}
{Spergel} D.,  et~al., 2015, arXiv e-prints, \href
  {https://ui.adsabs.harvard.edu/abs/2015arXiv150303757S} {p. arXiv:1503.03757}

\bibitem[\protect\citeauthoryear{{Stern} et~al.,}{{Stern}
  et~al.}{2018}]{stern2018}
{Stern} D.,  et~al., 2018, \mn@doi [\apj] {10.3847/1538-4357/aac726}, \href
  {http://adsabs.harvard.edu/abs/2018ApJ...864...27S} {864, 27}

\bibitem[\protect\citeauthoryear{{Tsai} et~al.,}{{Tsai} et~al.}{2015}]{tsai15}
{Tsai} C.-W.,  et~al., 2015, \mn@doi [\apj] {10.1088/0004-637X/805/2/90}, \href
  {http://adsabs.harvard.edu/abs/2015ApJ...805...90T} {805, 90}

\bibitem[\protect\citeauthoryear{{Wheelock} et~al.,}{{Wheelock}
  et~al.}{1994}]{iras}
{Wheelock} S.~L.,  et~al., 1994, NASA STI/Recon Technical Report N, \href
  {http://adsabs.harvard.edu/abs/1994STIN...9522539W} {95}

\bibitem[\protect\citeauthoryear{{Wright}, {Eisenhardt}, {Mainzer}, {Ressler},
  {Cutri}, {Jarrett}  \& {Kirkpatrick}}{{Wright} et~al.}{2010}]{wright10}
{Wright} E.,  {Eisenhardt} P.,  {Mainzer} A.,  {Ressler} M.~E.,  {Cutri} R.,
  {Jarrett} T.,   {Kirkpatrick} 2010, \mn@doi [\aj]
  {10.1088/0004-6256/140/6/1868}, \href
  {http://adsabs.harvard.edu/abs/2010AJ....140.1868W} {140, 1868}

\bibitem[\protect\citeauthoryear{{York} et~al.,}{{York} et~al.}{2000}]{sdss}
{York} D.~G.,  et~al., 2000, \mn@doi [\aj] {10.1086/301513}, \href
  {https://ui.adsabs.harvard.edu/abs/2000AJ....120.1579Y} {120, 1579}

\bibitem[\protect\citeauthoryear{{Zacharias}, {Finch}, {Girard}, {Henden},
  {Bartlett}, {Monet}  \& {Zacharias}}{{Zacharias} et~al.}{2013}]{ucac4}
{Zacharias} N.,  {Finch} C.~T.,  {Girard} T.~M.,  {Henden} A.,  {Bartlett}
  J.~L.,  {Monet} D.~G.,   {Zacharias} M.~I.,  2013, \mn@doi [\aj]
  {10.1088/0004-6256/145/2/44}, \href
  {https://ui.adsabs.harvard.edu/abs/2013AJ....145...44Z} {145, 44}

\makeatother
\end{thebibliography}

\appendix
\section{Complete unWISE Bitmask Details}
\label{app:bitmask}

This appendix provides full documentation of the mask bit meanings and methodology for our newly upgraded unWISE bitmask artifact flagging images.

\subsection{Bright source sample}
\label{sec:bright_sample}

A first step toward creating the unWISE bitmask images is defining a sample of sources which are sufficiently bright to potentially cause artifacts that require flagging. We maintain separate bright source catalogs in W1 and W2. In \cite{fulldepth_neo2}, we obtained these bright source samples via simple full-sky queries of the AllWISE catalog. Specifically, we selected all AllWISE sources with \verb|w1mpro| $<$ 9.5 (\verb|w2mpro| $<$ 8.3) in W1 (W2). We use Vega magnitudes and fluxes throughout this appendix unless otherwise noted. The equivalent thresholds in AB are 12.2 (11.64) in W1 (W2). Over the entire sky, this yields samples of 6,365,819 (2,221,020) sources in W1 (W2), with the density of such sources increasing dramatically near the Galactic plane. At $|b_{gal}| > 18^{\circ}$ (high enough Galactic latitude to fall within DESI's footprint), this bright source sample has on average $\sim$29 ($\sim$11) sources per deg$^2$ in W1 (W2).

Unfortunately, the AllWISE catalog lacks entries for some bright stars, and can report inaccurate (or ``null'') fluxes for some exceptionally bright objects that have far exceeded the WISE saturation threshold. To remedy this, the WISE team's ARTID artifact flagging module employed a custom ``Bright Source List'' that merged information from WISE itself, 2MASS, and even IRAS\footnote{\tiny{\url{http://wise2.ipac.caltech.edu/docs/release/allsky/expsup/sec4_4g.html\#bright_source_list}}}. Here, we take a similar approach to enhancing our AllWISE-based bright source list, but using only 2MASS $K_{S}$ magnitudes in a simplistic manner.

To merge 2MASS $K_S$ information into our AllWISE-based bright source list, we begin by selecting all 68,661 2MASS sources with $K_{S} < 5$. We then cross-match this 2MASS $K_S$ bright sample with our AllWISE sample, using a 15$''$ radius. This radius is chosen to allow for matching of stars with proper motions of up to $|\mu| \approx 1.5''$/yr given the $\sim$10 year AllWISE-2MASS time baseline. This proper motion threshold seems reasonable given that only $\sim$275 sources with $|\mu| > 1.5''$/yr are known, many of which (brown dwarfs, white dwarfs) are not sufficiently bright in W1 or W2 to require masking. For cases where an AllWISE bright source has a 2MASS $K_{S} < 5$ match within 15$''$, we replace our bright source catalog's \verb|w?mpro| magnitude with the 2MASS $K_{S}$ magnitude, if $K_{S} < $ \verb|w?mpro|. For 2MASS $K_S$ $<$ 5 sources with no AllWISE bright source match within 15$''$ or a match at $>$5$''$ separation, we instantiate a new object in our WISE bright source list at the 2MASS object's position and with its \verb|w?mpro| value set to the 2MASS $K_S$ magnitude. In W1 (W2), 4,582 (877) bright sources have their WISE magnitudes replaced with 2MASS $K_S$, and an additional 2,767 (2,414) WISE bright source sample entries are instantiated based on 2MASS.

In the future we will investigate ways to more accurately predict W1 and W2 magnitudes of bright stars based on multi-band 2MASS $JHK_{S}$ photometry, perhaps in combination with Gaia magnitudes and/or parallaxes. Also, given the availability of Gaia proper motions, improved 2MASS-AllWISE cross-matching could be enabled by propagating 2MASS positions to the AllWISE epoch. Lastly, unWISE coadds incorporate data spanning a considerable $\sim$8 year time period, so that one could imagine augmenting our WISE bright source list with motion information to instantiate multiple entries for bright sources that are moving very rapidly.

\subsection{PSF model thresholding}
\label{sec:psf_model}

\begin{figure*}
\begin{centering}
       \includegraphics[width=6.75in]{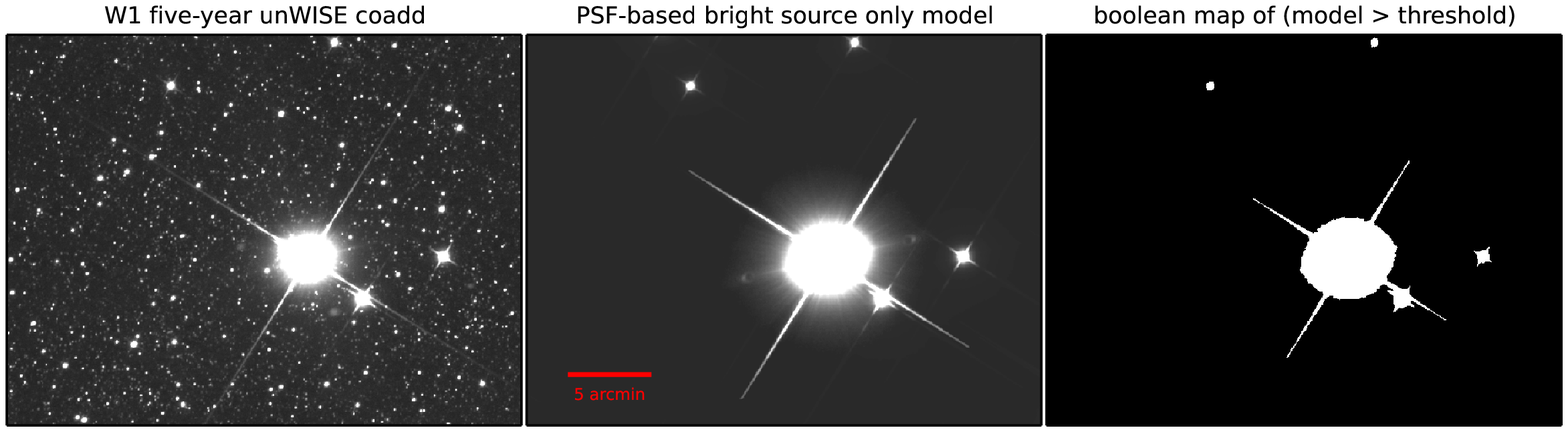}
      \caption{Schematic illustration of our PSF model thresholding procedure used in creating the unWISE bitmask images. Left: Portion of a five-year full-depth W1 unWISE coadd extracted from unWISE tile footprint 2415p106. The central, brightest source has AllWISE \texttt{w1mpro} $\approx$ 1.4. Middle: Bright source only model rendering of the same sky region, generated by combining the \citet{meisner14} W1 PSF model, the unWISE coadd WCS, and our bright source list described in $\S$\ref{sec:bright_sample}. Right: Bright source only model thresholded at 100 Vega nanomaggies per pixel to create a boolean bitmask, as we do when generating unWISE mask bits 0 and 1. The white regions represent pixels flagged as contaminated by bright sources.}
       \label{fig:psf_thresholding}
\end{centering}
\end{figure*}

\begin{figure*}
\begin{centering}
       \includegraphics[width=5in]{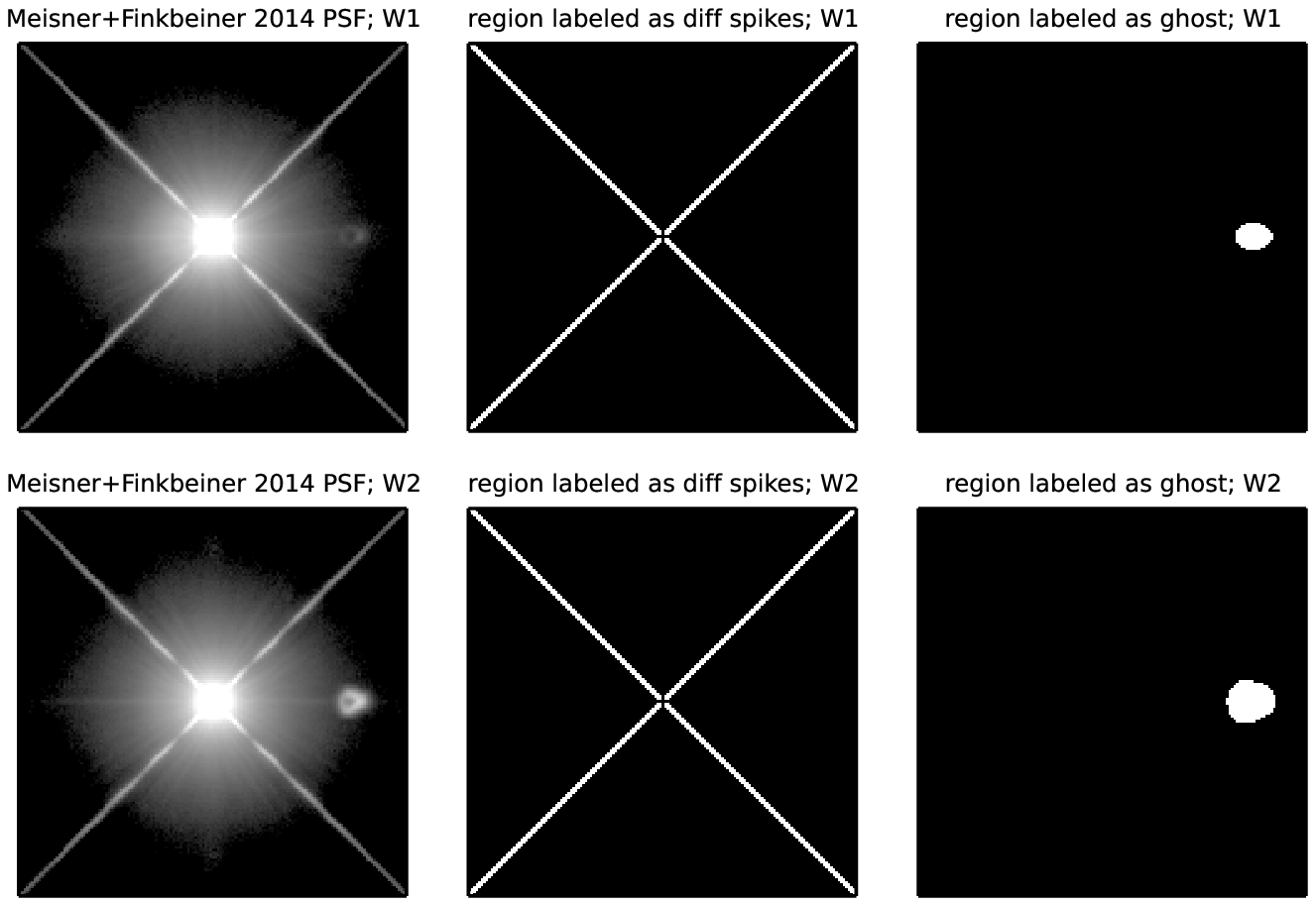}
      \caption{The PSF models used for creating our unWISE bitmasks, alongside corresponding boolean masks labeling PSF regions containing diffraction spikes and optical ghosts. Each subplot is 14.9$'$ on a side. Top left: W1 focal plane averaged PSF model. Top center: regions of the W1 PSF model labeled as containing diffraction spikes are shown in white on an otherwise black background. Top right: W1 PSF region labeled as being affected by an optical ghost. Bottom left: W2 focal plane averaged PSF model. Bottom center: regions of the W2 PSF model labeled as containing diffraction spikes. The top and bottom panels in the middle column are identical. Bottom right: W2 PSF region labeled as being affected by an optical ghost. The optical ghost is much more prominent in W2 than in W1.}
       \label{fig:psf_regions_labeled}
\end{centering}
\end{figure*}

Many of the unWISE mask bits are generated via a method that we refer to as ``PSF model thresholding'': on each tile's footprint in each band, we render a model image containing only objects from our bright source sample, then produce binary masks by flagging model image pixels brighter than a specified threshold as being ``contaminated''. Figure \ref{fig:psf_thresholding} illustrates an example of this methodology. The W1 and W2 PSF models themselves are shown in the left column of Figure \ref{fig:psf_regions_labeled}.

The key inputs to this procedure are the positions and fluxes from our bright source catalog ($\S$\ref{sec:bright_sample}), and the PSF models used to render the bright star profiles. For this purpose, we employ the \cite{meisner14} W1 and W2 PSF models. These PSF models natively exist in WISE detector coordinates, so we must perform several modifications before applying them to render models of unWISE coadds. The \cite{meisner14} PSFs include terms that account for PSF variation as a function of position within the detector. Because we wish to render models of unWISE coadd images that sample many detector positions at each bright source's location, we begin by constructing a PSF in each band that averages over detector position. We then use the WCS of each coadd footprint to rotate the averaged PSF such that it has an orientation matching that of the detector on the sky. Typically there are two such rotations for each unWISE tile footprint, corresponding to northward and southward WISE scans --- these two rotations of the PSF differ by 180$^{\circ}$ from one another. At very high ecliptic latitude, each bright star's location samples a substantial range of approach angles toward the ecliptic pole, meaning that the simple approximation of two discrete scan directions with PSF orientations separated by 180$^{\circ}$ no longer suffices. Currently, the unWISE mask bits constructed via PSF model thresholding do not account for this effect, which is only relevant over a very small fraction of the sky. With the PSF model averaged and rotated, for each bright source, we scale the PSF amplitude according to its magnitude listed in our bright source catalog and add it to the model image at the appropriate location. With the bright star model images in hand for each unWISE tile footprint, scan direction, and band we can proceed to apply various thresholding criteria, thereby flagging several types of artifacts. Full details of these thresholding steps are provided in subsequent sections.

One major limitation of the PSF model thresholding approach is that such flagging is limited in angular extent by the size of the PSF model. The \cite{meisner14} W1/W2 PSFs extend quite far into the wings, with sidelengths of $14.9'$ ($\sim$135 FWHM). Still, extremely bright stars can have diffraction spikes and ``halos'' reaching beyond the extent of our PSF models. Diffraction spikes become too long for the PSF models at parent magnitudes brighter than $\sim 4$. At $\sim$ 0$-$0.5 mag, the circularly symmetric component of the PSF profile begins to extend beyond our PSF model. Therefore, we also produce geometric masks for diffraction spikes ($\S$\ref{sec:geom_spike}) and bright source circular halos ($\S$\ref{sec:halo}), with radii that can extend far beyond the size of our PSF models.

\subsection{Bits 0-3: bright star `core and wings'}
\label{sec:p9}
Bits 0-3 are the original four mask bits from the \cite{fulldepth_neo2} unWISE bitmasks. They can be thought of as flagging the $\sim$0.6\% (0.4\%) of pixels most strongly affected by nearby bright source profiles at high Galactic latitude in W1 (W2). These masks were initially created to identify areas affected by difference imaging artifacts. Thus, these bits may provide sufficient bright star masking for unWISE-based difference imaging analyses, or other analyses in which the PSF modeling yields residuals comparably good to those from difference imaging.

Having rendered our bright source only model as described in $\S$\ref{sec:psf_model} for the relevant band, scan direction and unWISE tile footprint, generating mask bits 0-3 amounts to simply thresholding on the model pixel values. Based on examination of typical extragalactic sky regions, we adopt the same baseline threshold in W1 and W2: $\Gamma_0 = 100$ nanomaggies per pixel. We then scale the brightness threshold ($\Gamma$) toward larger values with increasing source density, in attempt to avoid masking an excessive fraction of area at low Galactic latitude. To implement this scaling of $\Gamma$ with source density, we begin by constructing a map of the number of Gaia DR2 \citep{gaia_dr2} sources per $N_{side}$ = 32 HEALPix pixel. We interpolate over the raw Gaia sources counts in HEALPix pixels affected by small-scale source density spikes (e.g., globular clusters) and localized dust clouds. We then interpolate off of the resulting source density map at coordinates corresponding to the center of the unWISE tile footprint for which we are constructing a bitmask, and refer to the resulting value as $n_{src}$ (units of Gaia sources per $N_{side} = 32$ HEALPix pixel). The threshold $\Gamma$ in nanomaggies/pixel as a function of source density is then determined by:

\begin{equation}
\label{equ:gamma}
\Gamma = \Gamma_0 + (max(n_{src}, 10^5) - 10^5) \times (\Gamma_{max} - \Gamma_0)/(n_{src, max}-10^5)
\end{equation}

So the threshold ramps linearly from its minimum value of $\Gamma_0$ to its peak value of $\Gamma_{max}$ between $n_{src} = 10^{5}$ and $n_{src} = n_{src, max}$. $\Gamma$ remains at its baseline value so long as $n_{src} < 10^{5}$, which is the case for $\sim$75\% of the sky. $n_{src, max}$ is $3.1 \times 10^6$ Gaia sources per $N_{side}$ = 32 HEALPix pixel. $\Gamma_{max}$ is 10,000 nanomaggies/pixel (9,250 nanomaggies/pixel) in W1 (W2).

Bit 0 (1) marks pixels which have values larger than $\Gamma$ in the southward (northward) scan W1 model rendering, as does bit 2 (3) for W2. In constructing each of bits 0-3, we also dilate the binary map of pixels above threshold by a $3 \times 3$ square kernel. We refer to these bits as `core and wings' because they capture the cores of bright star profiles, extending somewhat out into the wings so as to also capture diffraction spikes and optical ghosts for sufficiently bright objects. The most noticeable scan direction dependence seen in bits 0-3 pertains to the W2 ghost location, appearing on opposite sides of the parent bright star in opposite scan directions.

Note that bits 0-3 mask regions containing the parent bright source centroids themselves. Also, bits 0-3 account properly for cases in which some portion of a parent bright source's PSF profile overlaps the tile under consideration despite the parent's centroid falling outside of the tile boundaries.

\subsection{Bits 4-5: bright source saturation}
PSF modeling of bright sources will be compromised by attempting to fit saturated pixels near the profile core. To address this, unWISE mask bit 4 (5) marks ``saturated'' pixels in W1 (W2). Pixels flagged by bit 4 (5) are the subset of pixels flagged by bright source bits 0 OR 1 (2 OR 3) with model profile values which exceed thresholds of 85,000 (130,000) Vega nanomaggies in W1 (W2). Note that these per-band thresholds are fixed across the entire sky, with no spatial variation.

For the purposes of these unWISE mask bits, we have not intended to define the ``saturation'' threshold such that it truly corresponds to saturation of the WISE detectors. Instead, we are masking pixels which may be either fully saturated, potentially non-linear, or be bright enough to experience non-linearities to due interactions with outlier rejection during unWISE coaddition. At high Galactic latitude, the fraction of area masked by bit 4 (5) is $4.3 \times 10^{-5}$ ($3.0 \times 10^{-5}$).

\subsection{Bit 6: center of pixel not primary}
\label{sec:primary}
unWISE tile footprints are not mutually exclusive, and overlap by varying amounts depending on sky position. Typically, the overlap is $\sim$3$'$ along each boundary. As a result, one may sometimes wish to determine the ``best'' unWISE tile to consult for a specific (RA, Dec) sky location given the multiple tiles available to choose from. There are different possible ways to define ``best''. Here we define the best unWISE tile for a given (RA, Dec) as the tile which maximizes the minimum distance of that (RA, Dec) from any tile edge. For each pixel center's location within a given tile, one can compute whether this tile is the best tile for that pixel's (RA, Dec), or instead whether that sky location would be better analyzed in some other tile. Pixels marked with the ``center of pixel not primary'' unWISE mask bit are sky locations that would be better analyzed in another tile. The region flagged by this bit is a border along the edges of the tile, with a typical width of $\sim$35 pixels. 

\subsection{Bits 7-8: contamination from bright source with centroid off tile edge}
A bright source with PSF wings that contaminate an unWISE tile's footprint despite the bright source's centroid itself falling outside of the tile boundary can be problematic for image modeling analyses. For instance, the unWISE Catalog pipeline treats each unWISE tile footprint independently, detecting and modeling only those sources with centroids inside of the tile boundaries. As a result, the catalog creation process would by default be predisposed to model flux from the wings of off-edge bright sources as sums of large numbers of fainter point sources. To avoid this outcome, unWISE mask bits 7 (W1) and 8 (W2) flag pixels affected by bright stars with centroids outside of the tile's footprint. The unWISE Catalog pipeline recognizes these bits and uses them to preemptively make deblending less aggressive in the affected regions.

Bits 7-8 are based on bits 0-3 described in $\S$\ref{sec:p9}. For W1, if bit 0 and/or bit 1 is set and the parent bright star centroid is off the tile edge, then bit 7 is also set. For W2, if bit 2 and/or bit 3 is set and the parent bright star centroid is off the tile edge, then bit 8 is also set. Because the bit 7 (8) logic does not distinguish between bits 0-1 (2-3) in W1 (W2), information about scan direction is not retained in bits 7-8.

\subsection{Bit 9: resolved galaxy}
Certain image analyses may encounter problems in regions affected by resolved galaxies. For instance, the unWISE Catalog pipeline does not perform any galaxy model fitting, and by default deblends aggressively in an attempt to explain all flux above background as a sum of point sources. Flagging of resolved galaxies allows unWISE Catalog deblending to be made less aggressive in these regions; this avoids shredding resolved galaxies into large numbers of point sources. For our purposes, resolved galaxies are those with sizes $\gtrsim 6''$, which is the approximate W1/W2 PSF FWHM.

To flag resolved galaxies, we use an early version of the Legacy Survey Large Galaxy Atlas\footnote{\url{https://github.com/moustakas/LSLGA}. The LSLGA is based on HyperLeda \citep{hyperleda}.} (LSLGA). This catalog contains $\sim$2.1 million galaxies with angular sizes of $d_{25} \gtrsim 7''$, a size threshold coincidentally well-matched to the WISE FWHM. We visually inspected the largest (in terms of angular size) 300 LSLGA galaxies, manually removing a small number of SDSS \citep{sdss} filter edge reflection artifacts and very low surface brightness dwarf galaxies, and also tweaking a few of the $d_{25}$ parameters by eye. Based on this slightly modified LSLGA catalog, we use unWISE mask bit 9 to flag elliptical regions about each resolved galaxy, with the ellipse size set by $d_{25}$ and the appropriate shape/orientation determined by the axis ratio and position angle. $d_{25}$ is taken to be the major axis of the elliptical mask, as this visually appeared to work well. We mask circular regions in cases where position angles and/or axis ratios are not available. We also floor the $b/a$ axis ratio at 0.5 to avoid overly line-like masked regions. The median per-tile fraction of area flagged by the resolved galaxy bit is 0.08\%.

\subsection{Bit 10: big object}
\label{sec:big_obj}
This mask bit was inspired by (and largely copied from) the \verb|big_obj| mask bit within the \citet[SFD]{sfd98} dust map data products. Its purpose is to flag regions affected by the LMC, SMC and M31. For the LMC and SMC, the unWISE big object mask bit is set for each pixel based on querying the SFD \verb|big_obj| mask bit at that pixel center's (RA, Dec) coordinates. For M31, we use an elliptical mask with $a = 100'$, $a/b = 2.82$, and position angle of 35$^{\circ}$ east of north. This position  angle was chosen in order to best cover M31's outskirts.

\subsection{Bits 11-12 and 25-26: optical ghost}
\label{sec:ghost}

Bright source ghosts are well-captured by the \cite{meisner14} PSF models (see Figure \ref{fig:psf_regions_labeled}). As a result, we can employ our PSF model thresholding approach in order to specially flag regions affected by ghosts, making use of the thresholding criterion defined in $\S$\ref{sec:p9}. When rendering the same bright source only model used to create bits 0-3, we use the ghost regions labeled in the right column of Figure \ref{fig:psf_regions_labeled} to keep track of which pixels fall within bright source ghosts. We then generate our ghost-specific mask bits (11-12 in W2 and 25-26 in W1) by flagging those pixels that are within ghost-affected regions and  have model surface brightnesses larger than $\Gamma_{ghost} = 0.15\Gamma$, where $\Gamma$ is defined in Equation \ref{equ:gamma}. In other words, we apply a \textit{fainter} surface brightness threshold for masking within regions affected by ghosts, and assign the results of this modified thresholding to ghost-specific mask bits.

\subsection{Bits 13-20: latent}
\label{sec:latent}

\begin{figure*}
\begin{centering}
       \includegraphics[width=4in]{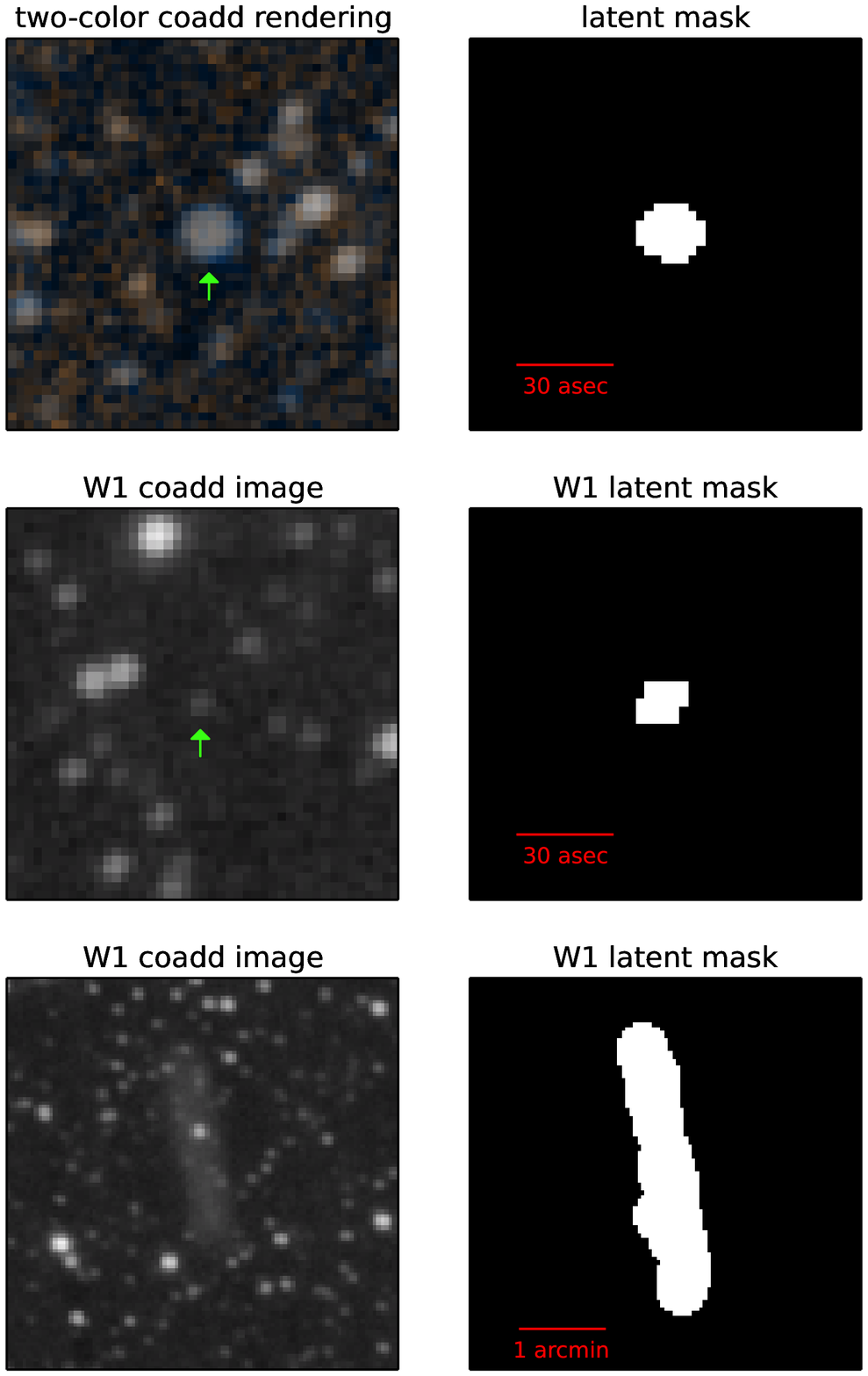}
      \caption{Examples of first order latents as imprinted on the unWISE coadds (left column) and flagged in our corresponding latent masks (right column). Top row: first latent of a $\sim 5$th magnitude parent source, pointed to by a green arrow. The two-color composite at left displays W1 as blue and W2 as orange. This example illustrates the slightly blue appearance of latents relative to sources with typical color of W1$-$W2 $\approx$ 0 (Vega). At right is a rendering of our unWISE bitmask in this latent's vicinity, with white pixels indicating that one or more of bits 13-16 is set. The location and size of the masked region match the observed artifact well. Middle row: W1 first latent of a \texttt{w1mpro} $\approx$ 8 parent source, which is close to the minimum brightness necessary to produce a latent. This latent covers a smaller region than does the first latent with $\sim5$th mag parent. Accordingly, a smaller region is flagged as affected by a W1 first latent, with white pixels at right indicating that bit 13 or 14 is set. Bottom row: latent arc produced by WISE scan direction variation at high absolute ecliptic latitude ($\beta \approx -76^{\circ}$). In this case the parent is a $\sim2.5$ mag star. The region flagged as affected by W1 first latents traces this arc, appearing to overflag somewhat as shown at right. Note that in the left column, the top panel is a two-color W1/W2 composite rendering, whereas the bottom two panels are W1-only grayscale renderings.} 

       \label{fig:first_latent_example}
\end{centering}
\end{figure*}

Persistence artifacts referred to as ``latents'' within the WISE documentation are present in W1 and W2 imaging. When a bright source is imaged and saturates one or more WISE pixels, the sky location imaged by those same detector pixels in subsequent exposures will display a diffuse blob-like latent feature. Latents are more pronounced in W1 than W2, and therefore tend to appear as somewhat blue smudges (see Figure \ref{fig:first_latent_example}). Exceptionally bright stars can cause latents that remain noticeable for many exposures while decaying in amplitude over time. We refer to the number of exposures since imaging of the parent bright source as the ``order'' of a latent. Currently, our unWISE masking accounts for only first and second order latents i.e., those arising one and two exposures subsequent to imaging of the parent bright source. Because WISE scans at $\sim$0.7$^{\circ}$ per exposure, the sky position of a latent is significantly offset from its parent source along the scan direction. This makes unflagged latents particularly troublesome for rare object searches --- latents are difficult to trace back to their parent bright sources by eye, and can also be misinterpreted as flux variable or moving objects because they appear at different sky locations during different WISE sky passes which have differing scan directions. Latent positions are deterministic, and could all be predicted exactly given the WCS and timestamp of every exposure in combination with a perfectly accurate bright source catalog.

As a first step toward creating image-level unWISE latent masks, we generate a catalog of latents appearing in each unWISE tile footprint and each band. To make the computation of latent positions within each coadd efficient, we precompute per-band lookup tables containing the MJDs and full WCS parameters of all $\sim$25 million single-exposure W1/W2 images. For each unWISE tile footprint in each band, we use the unWISE \verb|-frames| metadata table to determine the list of exposures contributing to the corresponding full-depth coadd. We then loop over these exposures, combining our WCS/MJD lookup table and bright source list ($\S$\ref{sec:bright_sample}) to determine the (RA, Dec) positions of latents appearing in each contributing exposure. We aggregate these latent world coordinates on a per-coadd basis, and convert this list of (RA, Dec) coordinates to pixel coordinates within the coadd under consideration. The result is one latent catalog per unWISE \verb|coadd_id| per band. In addition to world and coadd pixel coordinates, each such catalog contains a variety of metadata that will enable us to subsequently render image-level latent bitmasks. These metadata include, for each latent, the parent bright source magnitude, the WISE scan direction, and the latent's order (1 for first latent, 2 for second latent). In practice, when computing each coadd's catalog of latents, we do not analyze every single contributing exposure. Instead, we sort the contributing exposures by MJD and consider only every sixth exposure. This is done to reduce the computational cost of generating the latent catalog. Because a WISE sky pass typically includes $\gtrsim 12$ exposures at each sky location, and the WISE scan direction at a given sky position varies by $\lesssim0.5^{\circ}$ over the time period of six exposures, analyzing only the latents from every sixth exposure should not cause any image-level under-flagging downstream.

Only a subset of the objects within our bright source list are sufficiently bright to create latents. Table \ref{tab:latent_thresh} provides the magnitude thresholds that we employ for determining which bright sources are capable of causing first and second latents in each band. These thresholds are applied directly to the \verb|w?mpro| values in our bright source list, and are taken to be constant over time and across the entire sky.

\begin{table}
        \centering
        \caption{Latent parent magnitude thresholds (Vega).}
        \label{tab:latent_thresh}
        \begin{tabular}{lcc}
                \hline
                band & 1$^{st}$ latent threshold (mag) & 2$^{nd}$ latent threshold (mag) \\
                 \hline
                W1 & 8.3 & 6.2  \\
                W2 & 7.0 & 4.9  \\
                \hline
        \end{tabular}
\end{table}

To determine the angular size of the region that should be masked around each single-exposure latent location, we define an effective parent magnitude ($m_{eff, l}$) that takes into account several factors: the parent bright source's magnitude, the absolute ecliptic latitude and the background level. The following terms contribute to the determination of $m_{eff, l}$:

\begin{equation}
\label{equ:latent_bg_term}
\Delta_{bg, l} = f_{bg,l} \cdot log_{10}\Big[max(intmed/intmed_{0}, 1)\Big]
\end{equation}

\begin{equation}
\label{equ:delta_arc}
\Delta_{arc, l} = 2.5 \cdot log_{10}(max(\delta\theta, \delta\theta_0)/\delta\theta_0), \ \delta\theta_0 = 1^{\circ}
\end{equation}

The effective parent magnitude used to compute the size of each single-exposure latent is then:

\begin{equation}
\label{equ:latent_m_eff}
m_{eff, l} = m + \Delta_{bg, l} + \Delta_{arc, l}
\end{equation}

Where $m$ is the parent source's \verb|w?mpro| value taken from our bright source list. $intmed$ is the sky background level in WISE L1b DN. This sky background level is computed for each band on a per unWISE tile basis, by taking the median of the \verb|intmedian| values in the unWISE \verb|-frames| metadata table, restricted to frames that actually contributed to the coadd. The $intmed$ value used in Equation \ref{equ:latent_bg_term} is that of the unWISE tile for which the bitmask image is being generated. $intmed_{0}$ is a fiducial background level appropriate at high Galactic latitude and low ecliptic latitude. In Equation \ref{equ:latent_bg_term}, we adopt $intmed_0$ values of 25 DN (60 DN) in W1 (W2). The $f_{bg, l}$ prefactor is set to 2.38 (3.78) in W1 (W2), and $\Delta_{bg, l}$ is capped at 4.3 magnitudes. We use $\Delta_{bg, l}$ to account for the fact that an increased background level tends to reduce the region over which a latent's profile is non-negligible.

$\Delta_{arc, l}$ accounts for reduced latent surface brightness in the unWISE coadds due to increased arcing of latent imprints as $|\beta|$ becomes larger (see bottom row of Figure \ref{fig:first_latent_example}). Because of the continuous variation of the WISE scan direction at high ecliptic latitude, the coadd-level imprint of a bright source's multiple single-exposure latents is spread across an area which is $\sim \delta\theta/\delta\theta_0$ larger than it would have been in the ecliptic plane, with $\delta\theta$ given by:

\begin{equation}
\label{equ:dtheta}
\delta\theta = 0.78^{\circ}/cos(\beta)
\end{equation}

Equation \ref{equ:dtheta} can be thought of as the range of ecliptic longitude spanned by a WISE exposure (0.78$^{\circ}$ on a side), as a function of ecliptic latitude. This 
approximately corresponds to the range of ecliptic pole approach angles sampled as a function of $\beta$. $\delta\theta_0$ is set to $1^{\circ}$ because this is the approximate ecliptic pole approach angle spread within a given scan parity (northward or southward) for regions near the ecliptic plane. We cap $\delta\theta$ at $180^{\circ}$, at which point latents will have been spread out into complete rings surrounding their parent stars\footnote{This consideration is only relevant over a tiny portion of the sky, $|\beta| > 89.75^{\circ}$.}. Note that $\beta$ in Equation \ref{equ:dtheta} refers to the location of the parent bright source.

\begin{figure}
\begin{centering}
       \includegraphics[width=2.8in]{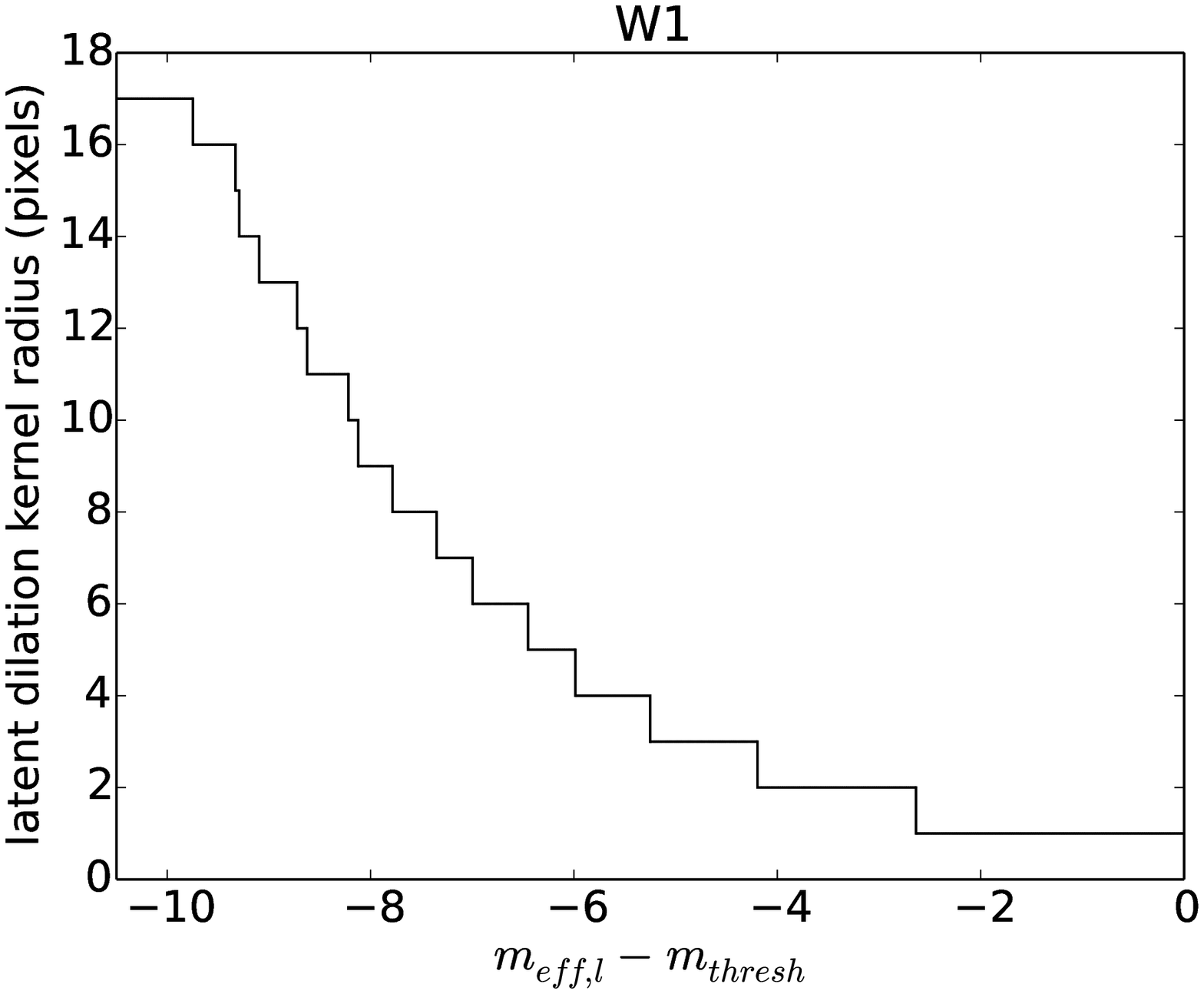}
      \caption{Radius of W1 per-latent circular binary dilation kernel as a function of difference between the effective parent magnitude of the latent (Equation \ref{equ:latent_m_eff}) and the latent parent magnitude threshold (Table \ref{tab:latent_thresh}). The corresponding plot in W2 is nearly identical. The radius is always floored at 1 pixel (corresponding to a $3 \times 3$ pixel rectangular dilation kernel), and capped at 17 pixels. For a given WISE band and order of latent, the region affected becomes larger as the parent's effective magnitude decreases. The increase of our adopted dilation radius toward lower $m_{eff,l}$ captures this behavior.}
       \label{fig:latent_radius}
\end{centering}
\end{figure}

To flag pixels affected by latents in a given unWISE coadd footprint, we begin by using nearest neighbor interpolation to mark each pixel corresponding to a single-exposure latent listed in the relevant latent catalog and having $m_{eff, l} \le m_{thresh}$, where $m_{thresh}$ is the magnitude threshold for the type of latent being considered (see Table \ref{tab:latent_thresh}). We then apply a binary dilation about each of these marked pixels using a circular kernel, with a radius that depends on the difference ($m_{eff, l} - m_{thresh}$). Figure \ref{fig:latent_radius} shows the dilation radius in pixels as a function of ($m_{eff, l} - m_{thresh}$). We determined the shape of this function by assuming that for a parent source of magnitude $m_{thresh}$, only the centermost pixel of the profile is sufficiently saturated to yield a non-negligible latent imprint. Then, using the \cite{meisner14} PSF model profile, we determined the radius of saturation that would result from making the total parent flux larger by a factor of 10$^{-(m_{eff, l} - m_{thresh})/2.5}$. 

As listed in Table \ref{tab:bitmask}, there are eight unWISE mask bits for latents, one for each possible combination of WISE band (W1 or W2), scan direction (north or south), and latent order (first or second).

\subsection{Bits 21-22: bright source centroid}
Many unWISE mask bits associated with bright sources flag the location of the parent bright object itself, a feature which may at times be considered undesirable. To address this situation, mask bit 21 (22) marks a 3 $\times$ 3 pixel box surrounding the centroid of each object in the W1 (W2) bright source sample described in $\S$\ref{sec:bright_sample}.

\subsection{Bits 23-24: AllWISE-like circular bright source halo}
\label{sec:halo}

In order to offer a set of unWISE mask bits similar to the AllWISE CC flags, we include bits marking circular ``halos'' around bright sources. Bit 23 (24) flags circular halos around bright sources in W1 (W2). Our halo radius formula, adapted from that of AllWISE, depends on the parent source's brightness, the sky background level, and absolute ecliptic latitude. The sky background level is relevant because the halos of bright stars become negligible at smaller radii when the sky background is higher, for example in the Galactic plane. The WISE coverage increases toward the ecliptic poles, so that for fixed parent star brightness and sky background level, the halo will tend to be appreciable relative to typical pixel noise out to a larger radius at higher absolute ecliptic latitude.

To construct our sample of halo parent sources, we downselect the bright star sample of $\S$\ref{sec:bright_sample} to sources with \verb|w?mpro| $<$ 8 in the relevant band\footnote{This requirement is an attempt to mirror the \texttt{w?mpro} $<$ 8 cut documented in item vi.2 of \scriptsize{\url{http://wise2.ipac.caltech.edu/docs/release/allsky/expsup/sec4_4g.html}}.}. We then use the following formulae to determine the halo radius, in arcseconds:

\begin{equation}
r_h = B \cdot 10^{a \cdot m_{eff, h} + b}
\end{equation}

\begin{equation}
\label{equ:halo_b}
B = c_{bg} \cdot log_{10}(intmed) + d_{bg}, \ B_{min} \le B \le B_{max}
\end{equation}

\begin{equation}
m_{eff, h} = m - 2.5 \cdot log_{10}\Big[\sqrt{1/max[cos(\beta), 0.2]}\Big]
\end{equation}

\begin{figure}
\begin{centering}
       \includegraphics[width=3.5in]{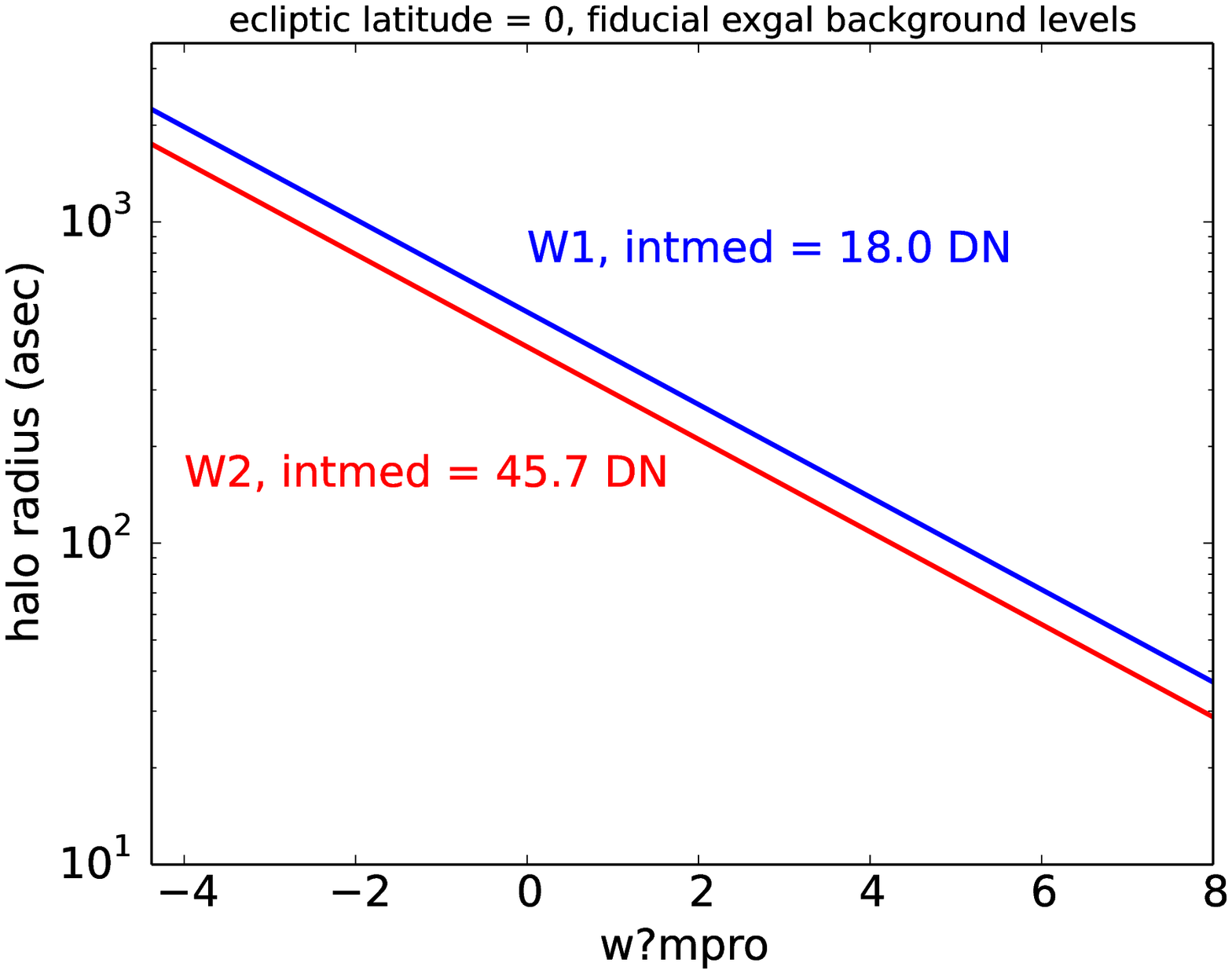}
      \caption{unWISE bitmask circular halo radii as a function of magnitude at $\beta$ = 0 and with fiducial high Galactic latitude sky background levels.}
       \label{fig:halo_radius_example}
\end{centering}
\end{figure}

The parameters of these equations are provided in Table \ref{tab:halo_params}. $m$ is the \verb|w1mpro| (\verb|w2mpro|) value from our bright source list for W1 (W2). For each halo parent source, the $intmed$ background level of the nearest tile center is adopted (see $\S$\ref{sec:latent} for a description of how per-tile $intmed$ values are calculated). $m_{eff,h}$ is an effective magnitude appropriate for use in the halo radius computation given the values of the other parameters. $m_{eff, h}$ is brighter than $m$ by an offset which tracks the signal-to-noise increase of a fixed flux source as its ecliptic latitude, and therefore WISE coverage, increases\footnote{The WISE frame coverage scales like 1/cos($\beta$), only deviating from this trend very nearby the ecliptic poles.}. The halo radius parameters were tuned based on inspection of five-year full-depth unWISE coadds --- different halo radii may be preferable when deeper or shallower unWISE coadds are being analyzed. Our halo radius functional forms and parameters are substantially rooted in those used by ARTID. Figure \ref{fig:halo_radius_example} shows our halo radii as a function of parent magnitude at $\beta$ = 0 and with fiducial high Galactic latitude sky background levels.

\begin{table}
        \centering
        \caption{Circular halo radius parameters.}
        \label{tab:halo_params}
        \begin{tabular}{lllllll}
                \hline
                band & $a$ & $b$ & $c_{bg}$ & $d_{bg}$ & $B_{min}$ & $B_{max}$\\
                 \hline
                W1 & $-$0.144 & 3.134 & $-$0.57 & 1.10 & 0.3 & 1.1 \\
                W2 & $-$0.144 & 3.134 & $-$1.35 & 2.30 & 0.3 & 1.1 \\
                \hline
        \end{tabular}
\end{table}

At high Galactic latitude ($|b_{gal}| > 18^{\circ}$), the typical fraction of area flagged by the unWISE circular halo masks is 1.2\% (0.7\%) in W1 (W2). We caution that circular masks are likely better implemented at the catalog level than at the image level, although image-level halo bitmasks do allow end users to conveniently dilate the halos as desired. Halos due to parent bright sources with centroids falling outside of a tile's boundaries but nevertheless partially impinging within that tile's footprint are always taken into account.

\subsection{Bits 27-28: PSF-based diffraction spike}
\label{sec:psf_spike}

Given the spatial extent of the PSF models we employ, we can use our PSF thresholding approach to make ``PSF-based'' diffraction spike masks out to a radius of up to $\sim$10.5$'$. The methodology for doing so is analogous to that described in $\S$\ref{sec:ghost} for generating ghost-specific mask bits. We track which pixels in our bright source only models are within diffraction spikes, based on the PSF regions labeled in the center column of Figure \ref{fig:psf_regions_labeled}. Within these diffraction spike regions, we apply a reduced brightness threshold of $\Gamma_{spike}$ = 0.05$\Gamma$, thereby generating mask bits specific to the diffraction spikes (bit 27 for W1 and 28 for W2). Because the WISE diffraction spikes are quite nearly symmetric under 180$^{\circ}$ rotation, the northward and southward scan direction diffraction spike masks are virtually identical. We therefore do not report PSF-based diffraction spikes using separate mask bits for each scan direction. Instead, each band's PSF-based diffraction spike bit is the OR of PSF-based diffraction spike flagging computed in the two scan directions.

\subsection{Bits 29-30: geometric diffraction spike}
\label{sec:geom_spike}
Diffraction spikes are captured to some extent in bits 0-3, and to a large extent in bits 27-28. However, for sufficiently bright stars, the diffraction spikes can reach beyond the boundaries of the \cite{meisner14} W1/W2 PSF models, which are limited to $\sim$15$'$ on a side. To handle such situations, we have implemented geometric diffraction spike mask bits for \verb|w?mpro| $\lesssim$ 6 sources, extending up to $\sim$1$^{\circ}$ from the parent's centroid.

Our geometric diffraction spike masks consist of straight lines emanating outward from the bright source centroid at angles of $45^{\circ}$, 135$^{\circ}$, 225$^{\circ}$, and 
$315^{\circ}$ from ecliptic north. In a manner similar to that used for determining circular halo size ($\S$\ref{sec:halo}), we consider multiple factors when calculating the geometric diffraction spike length: the parent source brightness, the sky background level, and the absolute ecliptic latitude. These factors are taken into account by computing an effective magnitude ($m_{eff, sp}$) for each object in our bright source sample. The following are terms that contribute to our computation of $m_{eff, sp}$:

\begin{equation}
\label{equ:spike_bg_penalty}
\Delta_{bg, sp} = 2.5 \cdot log_{10}\Big[max(intmed, intmed_0)/intmed_{0}\Big]
\end{equation}

\begin{equation}
\Delta_{cov, sp} = -2.5 \cdot log_{10}\Big[ \sqrt{1/cos(min(\beta, 80^{\circ}))}  \Big]
\end{equation}

The effective magnitude is then computed as:

\begin{equation}
\label{equ:m_eff_geom_spike}
m_{eff, sp} = m + \Delta_{bg, sp} + \Delta_{cov, sp} + \Delta_{fl, sp}
\end{equation}

Where $m$ is the \verb|w1mpro| (\verb|w2mpro|) value from our bright source list in W1 (W2). We only generate geometric diffraction spike masks for bright sources
with $m_{eff, sp} < 6$. $\Delta_{bg, sp}$ is a penalty that increases a bright source's magnitude (makes it considered to be effectively fainter) in regions of relatively high background. $intmed$ is the sky background level in WISE L1b DN, defined the same way as in our latent effective magnitude computation. When computing geometric diffraction spike radii, we floor $intmed$ at a fiducial extragalactic, low ecliptic latitude value $intmed_0$, taken to be 25 DN (60 DN) in W1 (W2). This forces $\Delta_{bg, sp}$ to be a strictly non-negative correction. $\Delta_{cov, sp}$ acts to decrease a bright source's effective magnitude (make it be considered brighter) with increasing absolute ecliptic latitude. At higher $|\beta|$, WISE provides larger frame coverage and hence reduced background noise in the unWISE coadds. The functional form of $\Delta_{bg, sp}$ results in a correction that tracks the decrease in background pixel noise with increasing $|\beta|$. In computing 
$\Delta_{bg, sp}$, we cap this ecliptic latitude correction at its $|\beta| = 80^{\circ}$ value, to avoid applying excessively large corrections very nearby the ecliptic poles.

At high ecliptic latitude, each unWISE coadd averages together frames with a significant spread in approach angles toward the ecliptic pole. Diffraction spikes therefore begin to take on a ``flared'' appearance at high $|\beta|$, and are ultimately washed out into nearly disk-like patterns in the immediate vicinity of the ecliptic poles. This results in decreased diffraction spike surface brightness at high ecliptic latitude, which we account for with the $\Delta_{fl, sp}$ term in Equation \ref{equ:m_eff_geom_spike}. The value of $\Delta_{fl, sp}$ is given by the right hand side of Equation \ref{equ:delta_arc}. In the case of geometric diffraction spikes, $\delta\theta$ (given by Equation \ref{equ:dtheta}) is capped at 90$^{\circ}$ because the flaring of diffraction spikes will form a complete ``disk'' around bright stars once the spread in ecliptic pole approach angles reaches this value (due to the fact that the single-exposure WISE diffraction spikes emanate outward at azimuthal angles spaced at 90$^{\circ}$ intervals).

Analytic functions of $m_{eff, sp}$ determine the geometric diffraction spike radius for each bright source. These functional forms and parameters are substantially rooted in those used by ARTID. The geometric diffraction spike radius is given by:

\begin{equation}
\label{equ:spike_radius}
r_{sp} = B_L \cdot 10^{a_L \cdot max(m_{eff,sp}, -2) + b_L} \cdot T(m_{eff, sp}) 
\end{equation}

\begin{equation}
\label{equ:taper}
T(m_{eff, sp}) = 1 - (6 - max(m_{eff, sp}, -2))/16
\end{equation}

\begin{figure}
\begin{centering}
       \includegraphics[width=3.5in]{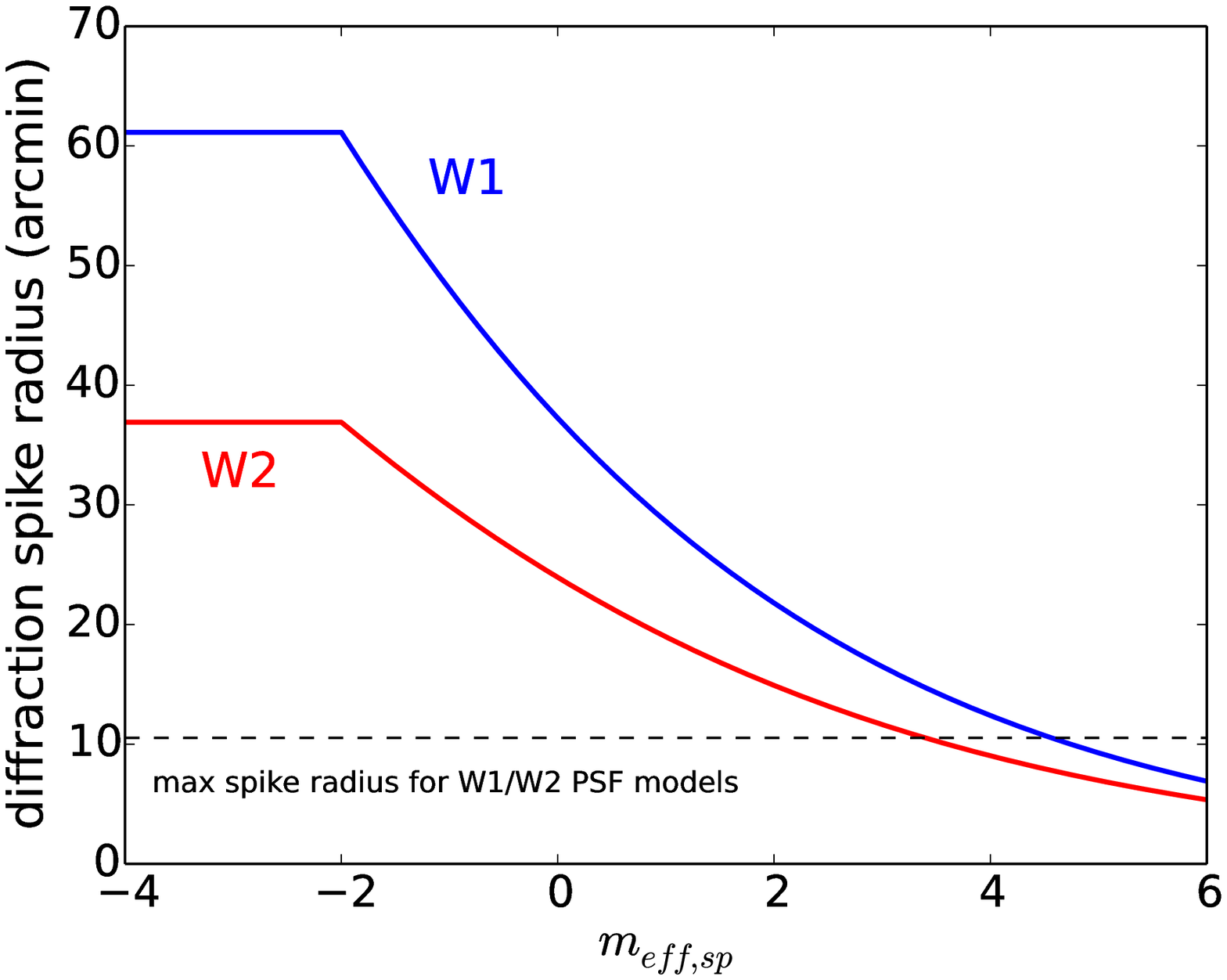}
      \caption{unWISE bitmask geometric diffraction spike radius as a function of effective magnitude, $m_{eff, sp}$. The blue (red) curve represents W1 (W2). The horizontal dashed black line is the maximum diffraction spike radius that can be accommodated within our PSF-based diffraction spike masking described in $\S$\ref{sec:psf_spike}. Our geometric diffraction spike masking goes into effect at $m_{eff, sp} = 6$, somewhat fainter than the effective magnitude at which the PSF-based diffraction spike masking can no longer be adequate ($m_{eff, sp} \approx 4$).}
       \label{fig:spike_radius}
\end{centering}
\end{figure}

$r_{sp}$ has units of arcseconds and $T(m_{eff, sp})$ is a tapering function that modulates the ARTID-like prescription in Equation \ref{equ:spike_radius}. Given that Equation \ref{equ:taper} floors $m_{eff, sp}$ at $-2$ and is only applied for $m_{eff,sp} < 6$, we always have $0.5 \le T(m_{eff, sp}) < 1$. $T(m_{eff, sp})$ ramps linearly with effective magnitude in the range $-2 \le m_{eff,sp} < 6$. The parameters of Equation \ref{equ:spike_radius} are listed in Table \ref{tab:spike_params}, and the geometric diffraction spike radius as a function of effective magnitude is shown in Figure \ref{fig:spike_radius}.

\begin{table}
        \centering
        \caption{Geometric diffraction spike radius parameters.}
        \label{tab:spike_params}
        \begin{tabular}{llll}
                \hline
                band & $B_L$ & $a_L$ & $b_L$ \\
                 \hline
                W1 & 7.07 & $-$0.195 & 3.38 \\
                W2 & 7.07 & $-$0.178 & 3.14 \\
                \hline
        \end{tabular}
\end{table}

We initially mark geometric diffraction spikes as narrow, $\sim$1 pixel wide lines of the appropriate orientation and length. We then dilate these narrow versions of the geometric diffraction spike masks by an amount which depends on the parent bright source effective magnitude, dilating more aggressively for brighter parent sources. Specifically, we dilate with a 5 $\times$ 5 pixel square kernel for $m_{eff, sp} > 0$, a 9 $\times$ 9 pixel square kernel for $-2 < m_{eff, sp}  \le 0$, and a 13 $\times$ 13 pixel square kernel for $m_{eff, sp} \le -2$.

Interior to each unWISE tile footprint, geometric diffraction spikes due to parent bright sources with centroids falling outside of the tile boundaries are taken into account. Although we reduce the geometric diffraction spike radii to account for flaring at high $|\beta|$, this flaring is not reflected in the morphology of the regions masked. We hope to implement this functionality in a future release of the unWISE bitmasks; this limitation of the present bitmasks is only relevant over a small fraction of the sky near the ecliptic poles.

\subsection{Collapsing scheme for DESI pre-imaging and the unWISE Catalog}
\label{sec:collapsing}

For some applications, the full content of the native unWISE bitmasks (Table \ref{tab:bitmask}) may not be necessary. In particular, the splitting of artifacts into multiple bits based on scan direction is only relevant for time domain applications. Our unWISE bitmasks are already being employed by catalogs which do not require scan direction dependent masking information --- the DESI pre-imaging surveys \citep{dey_overview} and the unWISE Catalog \citep{unwise_catalog}. For the purposes of these catalogs, we have developed a scheme to collapse the native 31-bit unWISE mask information into a smaller number of summary bits. The definitions of our 8 summary bits in terms of the native unWISE mask bits from Table \ref{tab:bitmask} are provided in Table \ref{tab:collapsing}.

\begin{table*}
        \centering
        \caption{Collapsing of unWISE mask bits for DESI pre-imaging and the unWISE Catalog.}
        \label{tab:collapsing}
        \begin{tabular}{llll}
                \hline
                Summary Bit & meaning & native unWISE & native unWISE\\
                 & & bit logic (W1) & bit logic (W2) \\
                 \hline
                0 &  bright star core and wings & 0 OR 1& 2 OR 3 \\
                1 & PSF-based diffraction spike & 27 & 28 \\
                2 & optical ghost & 25 OR 26 & 11 OR 12 \\
                3 & first latent & 13 OR 14 & 15 OR 16 \\
                4 & second latent & 17 OR 18 &  19 OR 20 \\
                5 & AllWISE-like circular halo & 23 & 24 \\
                6 & bright star saturation & 4 & 5 \\
                7 & geometric diffraction spike & 29 & 30 \\
                \hline
        \end{tabular}
\end{table*}

\subsection{Other possible future bitmask improvements}
\label{sec:improvements}
In the preceding sections, we have suggested a number of detailed unWISE bitmask improvements that we may implement in the future. In this section we propose a few additional upgrades that could be also be performed:

\begin{itemize}
  \item Homogenizing procedures for different mask bits --- More work could be done to homogenize the separate procedures used to create different unWISE mask bits. For example, our PSF thresholding scales back masking in crowded regions based on a map of Gaia source density, whereas the circular halo and latent mask bits do so based on sky background level.
  \item Overflagging in the Galactic plane --- Despite our efforts to limit masking in high source density regions, we still flag an excessively large fraction of area in some parts of the Galactic plane, mostly toward the Galactic center. Considering tiles that have centers within $1^{\circ}$ of the Galactic plane, those with 120$^{\circ} < l_{gal} < 240^{\circ}$ typically have $\sim$10-15\% of area masked. This fraction rises sharply toward the Galactic center, reaching $\sim$50\% at $(l_{gal}, b_{gal})= (\pm 45^{\circ}$, 0). At the Galactic center, the fraction of area masked peaks at $\sim$80-90\%. Users should indeed be cautious of essentially all unWISE measurements made near the Galactic center given the extreme crowding that results from the $\sim$6$''$ FWHM W1/W2 PSF. Nevertheless, further tuning of the ways in which unWISE masking is scaled based on source density and sky background level could be attempted.

  \item More finely pixelized bitmasks --- The unWISE bitmasks are pixelized at the native WISE pixel scale of 2.75$''$/pixel. Each such pixel covers an on-sky area equivalent to $> 100$ pixels of optical data from the Mosaic3 \citep{mosaic3} and Dark Energy Camera \citep{decam} instruments used for DESI pre-imaging. Thus, our current pixel size may limit the utility of applying our masks to WISE forced photometry based on detections in much higher resolution optical imaging. To address this, we could generate the unWISE bitmasks using a smaller pixel size.
  \item W3 and W4 --- In the future we could include information about W3 and W4 in our unWISE bitmask products.
  \item Nebulosity --- Nebulosity due to Galactic dust can be problematic for source detection and modeling analyses. The unWISE Catalog processing uses a neural 
 network to classify whether or not a sky region is affected by nebulosity \citep{decaps}, and reports this information via its \verb|flags_info| image product. This nebulosity flagging could potentially be added into our unWISE coadd bitmask images.
\end{itemize}

\end{document}